\documentclass[useAMS, usenatbib, a4paper]{mn2e}

\usepackage{mathrsfs}
\usepackage[space]{grffile}
\usepackage[T1]{fontenc}
\usepackage{aecompl}
\usepackage{times}
\usepackage{subfigure}
\usepackage{amsmath, amssymb}
\usepackage{amsfonts}
\usepackage{graphicx}
\usepackage{multirow}
\usepackage{hyperref}
\usepackage{url}
\usepackage{bm}
\usepackage{ulem} 
\usepackage[abs]{overpic}
\usepackage{todonotes}

\renewcommand{\d}[1]{\ensuremath{\operatorname{d}\!{#1}}}
\newcommand{\vect}[1]{\mbox{\boldmath ${#1}$}}
\newcommand {\apgt} {\ {\raise-.5ex\hbox{$\buildrel>\over\sim$}}\ }
\newcommand {\aplt} {\ {\raise-.5ex\hbox{$\buildrel<\over\sim$}}\ }

\def\myputfigure#1#2#3#4#5%
{\vskip#5pt\makebox[0pt]{\hskip#2in
\includegraphics[width=#3\textwidth]{#1}}\vskip#4pt\hfill}

\newcommand{\hi}{\mathrm{H}\textsc{i} }
\newcommand{\xh}{x_{\mathrm{H}\textsc{i}} }

\title[Fast estimator for the bispectrum \& beyond.]{A fast estimator for
the bispectrum and beyond -
A practical method for measuring non-Gaussianity in 21-cm maps.}
\author[C. A. Watkinson, S. Majumdar \& J. R. Pritchard]
{Catherine ~A.~Watkinson$^1$\thanks{Email: \href{mailto:catherine.watkinson@gmail.com}
{\protect\nolinkurl{catherine.watkinson@gmail.com}}}, Suman Majumdar$^2$, Jonathan ~R.~Pritchard$^2$ \& Rajesh Mondal$^3$\\
$^1$Department of Physics \& Astronomy, UCL, Gower Street, London, WC1E 6BT \\
$^2$Department of Physics, Blackett Laboratory, Imperial College, London, SW7 2AZ, UK \\
$^3$Department of Physics \& Centre for Theoretical Studies, Indian Institute of Technology Kharagpur,
Kharagpur - 721302, India \\}

\date{\today}

\pubyear{\number\year}

\begin{document}
\maketitle

\begin{abstract}
In this paper we establish the accuracy and robustness of a fast estimator
for the bispectrum - the `FFT bispectrum estimator'.
The implementation of the estimator presented here
offers speed and simplicity benefits over a direct-measurement approach.
We also generalise the derivation so it may be easily be applied to any order
polyspectra, such as the trispectrum, with the cost of only a handful of FFTs.
All lower order statistics can also be calculated simultaneously for little extra cost.
To test the estimator we make use of a non-linear density field, and for a more
strongly non-Gaussian test case we use a toy-model of reionization in which ionized bubbles
at a given redshift are all of equal size and are randomly distributed.
Our tests find that the FFT estimator remains accurate over a wide range of $k$,
and so should be extremely useful for analysis of 21-cm observations.
The speed of the FFT bispectrum estimator
makes it suitable for sampling applications,
such as Bayesian inference.
The algorithm we describe should prove valuable in the analysis of simulations
and observations, and whilst we apply it within the field of cosmology,
this estimator is useful in any field that deals with non-Gaussian data.
\end{abstract}
\begin{keywords}
methods: statistical -- dark ages, reionization, first stars -- intergalactic medium -- cosmology: theory.
\end{keywords}

\section{Introduction}\label{sec:intro}
The first stars and galaxies produced copious amounts of
UV radiation, which was capable of ionizing neutral hydrogen. The short mean free
path of this radiation means that well defined ionized bubbles form and grow around
sources, eventually merging to complete the reionization of the Universe.
This phase change of the Universe's hydrogen content, from neutral to ionized,
is known as the Epoch of Reionization (EoR).
We refer the interested reader to \citealt{Loeb2013} and \citealt{Pritchard2011}
for an overview of reionization.
The resulting distribution of neutral hydrogen is expected
to be extremely non-Gaussian, for example \citet{Harker2009},  \cite{Friedrich2010},
\citet{Watkinson2014}, \citet{Dixon2015a}, \cite{Mondal2016}, and \cite{Kakiichi2017}.

Atomic hydrogen may emit or absorb radiation with a $\lambda\sim 21$ cm (at rest) due
to an hyperfine transition in its lowest energy level, which is caused by the magnetic moment
of the bound electron flipping relative to the proton nucleus \citep{Field1958,Field1959a}.
Several existing radio telescopes (e.g. LOFAR\footnote{The LOw Frequency ARray \url{http://www.lofar.org/}},
PAPER\footnote{The Precision Array to Probe Epoch of Reionization
\url{http://eor.berkeley.edu/}} and MWA\footnote{The Murchison Wide-field Array
\url{http://www.mwatelescope.org/}}), and
future radio telescopes (e.g. HERA\footnote{The Hydrogen Epoch of Reionization Array
\url{http://reionization.org/}} and the
SKA\footnote{Square Kilometre Array \url{https://www.skatelescope.org}}),
are aiming to detect fluctuations in this 21-cm signal from the high-redshift Universe
\citep{Mellema2013, Ali2015, Beardsley2016b, DeBoer2017, Patil2017a}.
To complement this effort, there are also experiments seeking to measure
the average (or global) 21-cm signal, such as
EDGES\footnote{The Experiment to Detect the Global EoR Signal
\url{http://www.haystack.mit.edu/ast/arrays/Edges/}}, SARAS,
and
DARE\footnote{The Dark Ages Radio Explorer \url{http://lunar.colorado.edu/dare/}}
\citep{bowman2010, Burns2012, Singh2017}.
With such observations, we hope to learn about the process of reionization,
and the nature of the first generations of stars and galaxies.

The lowest order statistic that is sensitive to non-Gaussianity in a dataset
is the three-point correlation function, i.e. the excess probability as a function of three
points in the dataset.
The Fourier equivalent of the three-point correlation function is
the bispectrum, defined by
\begin{equation}
\begin{split}
(2\pi)^3 B(\vect{k}_1, \vect{k}_2, \vect{k}_3) \delta^{\mathrm{D}}(\vect{k}_1 + \vect{k}_2 +  \vect{k}_3  )=
\langle \Delta(\vect{k}_1)\Delta(\vect{k}_2)\Delta(\vect{k}_3)\rangle \,,\\\label{eq:bi_definition}
\end{split}
\end{equation}
where angular brackets describe an ensemble-averaged quantity,
and $\Delta(\vect{k})$ is the Fourier Transform of the density contrast field
$\delta(\vect{x}) = \rho(\vect{x})/\langle\rho(\vect{x}) \rangle - 1$.
The bispectrum has been studied extensively to constrain non-Gaussianity
in large-scale structure, see for example analysis of BOSS data by \citet{Gil-Marin2016},
and the cosmic microwave background \citep{PlanckCollaboration2015b}.

The skewness{\footnote{The skewness $\gamma$ measures the asymmetry of the data's
probability density function, i.e. $\gamma = \langle (x_i-\overline{x})^3\rangle$
(where $N$ describes the total pixels, and $\overline{x}$ the mean of the
pixel values $x_i$), and is usually normalised
by the cube of the standard deviation $\sigma^3$.}}
is the zero-separation 3-point correlation function
$\xi(\vect{x}_1, \vect{x}_2, \vect{x}_3)$, which is related to the bispectrum
$B(\vect{k}_1, \vect{k}_2, \vect{k}_3)$ (where
$\vect{k}_3 = - \vect{k}_1 - \vect{k}_2$)
as,

\begin{equation}
\begin{split}
\gamma = \xi(0, 0, 0) = \int \frac{\mathrm{d}^3\mathrm{k}_1}{(2\pi)^3}
\int \frac{\mathrm{d}^3\mathrm{k}_2}{(2\pi)^3} B(\vect{k}_1, \vect{k}_2, \vect{k}_3)
\,. \label{eqn:skew}\\
\end{split}
\end{equation}
Studies of the skewness of 21-cm simulated maps
have highlighted that there is a great deal of information to be gained from
moving beyond the power spectrum, which to date has been the main focus of high-$z$ 21-cm
studies \citep{Harker2009, Watkinson2014, Shimabukuro2014, Watkinson2015, Watkinson2015a}.

As the bispectrum is a function of both the size and shape of triangles formed by
a closed loop of $k$-vectors,
there will be more information to be gained by measuring the
bispectrum from 21-cm maps than there is from measuring only the skewness
\citep{Shimabukuro2015, Shimabukuro2016a}.
The challenge we face is that there is a huge choice of triangle configurations
that may be considered; furthermore, the statistic is very time
consuming to evaluate, typically involving a nested loop through a Fourier transformed (FT)
box\footnote{For a real field $V({\bf x})$, which satisfies the
Hermitian condition $V^{*}({\bf k}) = V(-{\bf k})$,
only half the FT box need be looped through.}
in order to evaluate the bispectrum using direct measurement,

\begin{equation}
\begin{split}
B(\vect{k}_1, \vect{k}_2, \vect{k}_3)= \frac{1}{(2\pi)^3}
\frac{1}{N_{\mathrm{tri}}}\sum\limits_{m\in\mathrm{Tri}_{123}}
\Delta(\vect{k}_1)\Delta(\vect{k}_2)\Delta(\vect{k}_3)
\,.\label{eq:bi_direct}
\end{split}
\end{equation}
$\mathrm{Tri}_{123}$ describes the set
of $\{ \vect{k_1}, \vect{k_2}, \vect{k_3} \}$
which form a triangle, i.e where $\vect{k_1} + \vect{k_2} + \vect{k_3}=0$.

Simulations and observations of the high-$z$ 21-cm signal produce
large datasets, for example the SKA will have of order 20,000 pixels per
frequency slice,\footnote{This calculation is based on the SKA 2015 configuration,
document number SKA-TEL-SKO-0000308
\url{http://skatelescope.org/wp-content/uploads/2014/03/SKA-TEL-SKO-0000308_SKA1_System_Baseline_v2_DescriptionRev01-part-1-signed.pdf} }
and a typical simulation contains $>500^3$ pixels.
It will therefore be very time consuming to calculate the above.  In
order to make bispectrum studies more tractable, we investigate a more
efficient estimator, which we call the `FFT bispectrum estimator'.
This estimator is a recasting of Equation \ref{eq:bi_direct} that
allows the bispectrum to be calculated with a single loop through the
FT dataset, followed by six Fast-Fourier Transforms (FFT) and a loop
through the real-space data.  Importantly, it is trivial to extend
this estimator to higher orders than three, we therefore present the
general form of the estimator that may be used to calculate an
$p^{\mathrm{th}}$-order statistic or polyspectrum.

This approach for measuring the bispectrum is described in
\citet{Scoccimarro2015} and \citet{Sefusatti2015}.
The technique has been used to measure the bispectrum from density fields and
galaxy clustering, initially without mention, for example \citet{Scoccimarro2000},
\citet{Feldman2001}, and \citet{Scoccimarro2001}. More recently, it
has been explicitly applied;
for example, \citet{Regan2011}, \citet{Schmittfull2012}, \citet{Schneider2016},
\citet{Gil-Marin2016}, and \citet{Byun2017}.
A similar approach has also been applied using spherical
harmonic transforms, instead of FFTs, for CMB data in \citet{Komatsu2001}.
A similar technique has also been used to speed up calculations of the three-point
correlation function \citep{Slepian2015a}.
The aim of this work is (1) to describe how the estimator may practically be calculated,
and (2) to test the performance of the FFT polyspectra estimator as applied to the
bispectrum and power spectrum in the context of 21-cm cosmology,
comparing it to both theoretical predictions and a direct-measurement method.
It is also hoped that this paper, by devoting full attention to the practical
application of the FFT estimator, will raise the attention of the 21-cm community
(as well as other research communities) to its existence.

This paper is structured as follows: In Section \ref{sec:Nspec}, we
present the derivation of the FFT polyspectrum estimator, and discuss
some nice properties of this approach for measuring polyspectra with
$p>2$. We also describe an algorithm that efficiently applies this approach.
We then specialise, in Section \ref{sec:Test}, to the case of
the bispectrum in order to test the effectiveness of the FFT
polyspectrum estimator.  We measure the bispectrum from a non-linearly
evolved density field to evaluate the estimator's accuracy on a weakly
non-Gaussian dataset.  We then use a toy model for reionization to
test the FFT estimator's accuracy when measuring the bispectrum from a
strongly non-Gaussian dataset.  Finally, in Section \ref{sec:Conc} we
conclude the findings of this work.  Unless otherwise stated, all
units are comoving.

\section{The FFT polyspectrum estimator}\label{sec:Nspec}
In this section we expand on a derivation in the thesis of
\citet{Jeong2010a}, which in turn builds on the thesis of
\citet{Sefusatti2005},
to present a general expression for estimating the
$p^{\mathrm{th}}$-order polyspectrum utilising FFTs.
We also describe an algorithm that applies this method for measuring polyspectra.
We will then
specialise to the case of $p=2$ (the power spectrum) and $p=3$ (the
bispectrum).
We will use the following FFT conventions for the remains
of this paper,

\begin{equation}
\begin{split}
\delta(\vect{x})&=\frac{1}{V}\sum \Delta(\vect{k})\mathrm{e}^{i\vect{k}\cdot\vect{x}}\,,\\
\Delta(\vect{k})&=H\sum \delta(\vect{x})\mathrm{e}^{-i\vect{k}\cdot\vect{x}}\,,
\label{eq:FFTconventions}
\end{split}
\end{equation}
\noindent where $H = V/N_{\mathrm{pix}}$, $V$ is the volume under
analysis, and $N_{\mathrm{pix}}$ is the total number of pixels in that
volume.

As our simulations and data will be pixelised it is useful to write
the polyspectrum estimator in terms of dimensionless pixel
co-ordinates, translating $\vect{k} = k_{\textsc{f}}\vect{m}$, where
$\vect{m}$ is a dimensionless integer triplet $(m_x, m_y, m_z)$ and
$k_{\textsc{f}} = 2\pi/L$ where $L$ is the simulated box length on a
side \footnote{If we were working with non-cubic data then
  $\boldsymbol{n} = (x/L_x, y/L_y, z/L_z)$ and $\vect{m} =
  (k_x\,L_x/(2\pi), k_y\,L_y/(2\pi), k_z\,L_z/(2\pi)$ where $L_i$ is
  the length of box side in the $i$ axis.  However, for the sake of
  simplicity our derivation is formulated for a cube for which each
  side is the same length, were this not the case there would
  technically be a different fundamental $k_{\textsc{f}}$ for each
  axis. Regardless, this factor reduces to $1/V$ in the final
  estimator which is calculated in the same way regardless of whether
  the data volume is cubic or not.}.  The delta function has
properties such that we may write $\delta^{\mathrm{D}}[a\,\vect{x}] =
\prod\limits_j|a|^{-1}\delta^{\mathrm{D}}(x_j)$, where $j$ describes
the components that make up the vector $\vect{x}$, and $a$ is a
non-zero scalar.  We can therefore rewrite the Dirac delta function in
dimensionless pixel co-ordinates $(m_x, m_y, m_z)$ as,
\begin{equation}
\begin{split}
\delta^{\mathrm{D}}(\vect{k})&=\delta^{\mathrm{D}}(k_{\textsc{f}}\vect{m})\,,\\
&=\prod_j\delta^{\mathrm{D}}(k_{\textsc{f}}m_j) = \prod_j\frac{1}{k_{\textsc{f}}}\delta^{\mathrm{D}}(m_j)\,,\\
&=\prod_j\frac{1}{k_{\textsc{f}}}\delta^{\mathrm{K}}(m_j) = \frac{1}{k_{\textsc{f}}^3}\delta^{\mathrm{K}}(\vect{m})\,.
\end{split}
\end{equation}
As our dataset is discrete, we have converted to the Kronecker-delta
function $\delta^{\mathrm{K}}(m_j)$, the discrete realisation of the
Dirac-delta function, in the last line.  We also need to connect the
unnormalised output of the FFTW algorithm
$\Delta_{\textsc{fft}}(\vect{k})$ to the theoretical
$\Delta(\vect{k})$ as described in Equations \ref{eq:bi_direct} and
\ref{eq:FFTconventions},
\begin{equation}
\begin{split}
\Delta_{\textsc{fft}}(\vect{m}) &=
\sum\limits_{\boldsymbol{r}}\delta(\vect{x})\,\mathrm{e}^{-i\boldsymbol{x}\cdot\boldsymbol{k}}
=\frac{\Delta(\vect{k})}{H}\,,\\
&=\sum\limits_{\boldsymbol{n}}\delta(\vect{n})\,\mathrm{e}^{-i2\pi\boldsymbol{m} \cdot \boldsymbol{n}/N_{\mathrm{side}}}\,,\\
\label{eq:DeltaFFTtoDelta}
\end{split}
\end{equation}
where $N_{\mathrm{side}}$ is the number of pixels on each side of the cube,
and spatial co-ordinates are related to pixel co-ordinates as
$\vect{x} = \vect{n}\,L/N_{\mathrm{side}}$.
With these conversions in hand we can write down an expression
for the polyspectrum as measured from a discrete dataset,
$\mathscr{P}(\vect{k}_1, \vect{k}_2,\, ...\, \vect{k}_p)$,

\begin{equation}
\begin{split}
(2\pi)^3 \mathscr{P}(\vect{k}_1, \vect{k}_2,\, ...\, \vect{k}_p)
\delta^{\mathrm{D}}(\vect{k}_1 + \vect{k}_2 &\, ... \, +  \vect{k}_p  )\\
&= \left\langle \prod\limits_i^p\Delta(\vect{k}_i)\right\rangle \,,\\
(2\pi)^3 \mathscr{P}(\vect{k}_1, \vect{k}_2,\, ...\, \vect{k}_p)
\delta^{\mathrm{K}}(\vect{k}_1 + \vect{k}_2& \, ... \, +  \vect{k}_p  )\\
&\approx H^p\left\langle \prod\limits_i^p\Delta_{\textsc{fft}}(k_{\textsc{f}}\vect{m}_i)\right\rangle \,,\\
\label{eq:gen_Nspec1}
\end{split}
\end{equation}
where we implement the conversion to discrete Kronecker delta function
and unnormalised FFTW $\Delta_{\textsc{fft}}(\vect{m})$ in the second
line.
Because our dataset is discrete, we are forced to work with a bin width
of at least $k_{\textsc{f}}$, the RHS therefore becomes an approximation of the LHS.
Cancellations, and enforcing the delta function on the left then
gives us,
\begin{equation}
\begin{split}
\mathscr{P}(\vect{k}_1, \vect{k}_2,\, ...\, \vect{k}_p)
&\approx H^p \frac{1}{V} \\
& \times\left\langle \prod\limits_i^p \delta^{\mathrm{K}}(\vect{m}_1 + \vect{m}_2 +\, ...
\, +  \vect{m}_p  )\,\Delta_{\textsc{fft}}(\vect{m}_i)\right\rangle  \,.\\
\label{eq:gen_Nspec2}
\end{split}
\end{equation}
We can also incorporate an arbitrary bin width $s$ such that,
\begin{equation}
\begin{split}
&\mathscr{P}(\vect{k}_1, \vect{k}_2,\, ...\, \vect{k}_p)
\approx H^p \frac{1}{V} \,\frac{1}{N_{\mathrm{poly}}}\,\\
&\times \sum\limits_{\boldsymbol{l}_1\pm s/2}\,...\, 
\sum\limits_{\boldsymbol{l}_p\pm s/2} \prod\limits_i^p
\delta^{\mathrm{K}}(\vect{m}_1 + \vect{m}_2\, ... \, +  \vect{m}_p  )\,\Delta_{\textsc{fft}}(\vect{m}_i)\,,  \\
&= H^p \frac{1}{V}\\
&\times \frac{
\sum\limits_{\boldsymbol{l}_1\pm s/2}\,...\,
\sum\limits_{\boldsymbol{l}_p\pm s/2} \prod\limits_i^p
\delta^{\mathrm{K}}(\vect{m}_1 + \vect{m}_2\, ... \, +  \vect{m}_p)\Delta_{\textsc{fft}}(\boldsymbol{m}_i)}
{\sum\limits_{\boldsymbol{l}_1\pm s/2}\,...\,
\sum\limits_{\boldsymbol{l}_p\pm s/2}
\delta^{\mathrm{K}}(\vect{m}_1 + \vect{m}_2\, ... \, +  \vect{m}_p)}\,,\\
\label{eq:gen_Nspec3}
\end{split}
\end{equation}
where $\boldsymbol{l}_i = |(\boldsymbol{k}_i/k_{\textsc{f}})-\vect{m_i}|$
and the sums are over all $\vect{m_i}$ vectors that
fall within a bin width of $\vect{k}_i/k_{\textsc{f}}$, i.e. all $k$-space
pixels for which $\boldsymbol{l}_i \le s/2$.
$N_{\mathrm{poly}}$ is the number of polygons formed by
$\vect{m}_1 + \vect{m}_2 \, ... \, + \vect{m}_p =
0$.
Whilst it is possible to use any value for $s$ within this framework,
we advise that the binwidth is kept to that of a pixel.
$N_{\mathrm{poly}}$ can be written in terms of a sum over the
Kronecker delta function when modes meet the above requirements, as
per the last line of Equation \ref{eq:gen_Nspec3}.

Recalling that $\vect{x} = \vect{n}\,L/N_{\mathrm{side}}$,
the Kronecker delta may be written as,
\begin{equation}
\begin{split}
&\delta^{\mathrm{K}}(\vect{m}_1 + \vect{m}_2 \,...\,+ \vect{m}_p )\,, \\
&= \frac{1}{N_{\mathrm{pix}}}\sum\limits_{\boldsymbol{n}}^{N_{\mathrm{pix}}}\,
\mathrm{e}^{i2\pi\vect{n}\cdot(\vect{m}_1 + \vect{m}_2 \,...\, + \vect{m}_p )/N_{\mathrm{side}}}\,, \\
&= \frac{1}{N_\mathrm{pix}}\sum\limits_{\boldsymbol{n}}^{N_{\mathrm{pix}}}\,
\prod\limits_i^p\mathrm{e}^{i2\pi \boldsymbol{n}\cdot \boldsymbol{m}_i/N_{\mathrm{side}}}\,.\\
\end{split}
\end{equation}
Equation \ref{eq:gen_Nspec3} then becomes,
\begin{equation}
\begin{split}
&\mathscr{P}(\vect{k}_1, \vect{k}_2,\, ...\, \vect{k}_p)
\approx H^p \frac{1}{V}\\
&\times \frac{
\sum\limits_{\boldsymbol{n}}^{N_{\mathrm{pix}}}\,
\left[\sum\limits_{\boldsymbol{l}_1\pm s/2}
\,...\,
\sum\limits_{\boldsymbol{l}_p\pm s/2}
\prod\limits_i^p \Delta_{\textsc{fft}}(\boldsymbol{m}_i)\mathrm{e}^{i2\pi
\boldsymbol{n}\cdot \boldsymbol{m}_i/N_{\mathrm{side}}} \right]}
{\sum\limits_{\boldsymbol{n}}^{N_{\mathrm{pix}}}\,
\left[\sum\limits_{\boldsymbol{l}_1\pm s/2}
\,...\,
\sum\limits_{\boldsymbol{l}_p\pm s/2}
\prod\limits_i^p\mathrm{e}^{i2\pi \boldsymbol{n}\cdot
\boldsymbol{m}_i/N_{\mathrm{side}}} \right]}\,.\\
\label{eq:gen_Nspec4}
\end{split}
\end{equation}
To modularise the calculation we define the following,
\begin{equation}
\begin{split}
\delta(\boldsymbol{n},\,\boldsymbol{k}_i) &=
\sum\limits_{\boldsymbol{l}_i\pm s/2}
\Delta_{\textsc{fft}}(\boldsymbol{m}_i)\mathrm{e}^{i2\pi \boldsymbol{n}
\cdot \boldsymbol{m}_i/N_{\mathrm{side}}}\,,\\
I(\boldsymbol{n},\,\boldsymbol{k}_i)
&= \sum\limits_{\boldsymbol{l}_i\pm s/2}
\mathrm{e}^{i2\pi \boldsymbol{n}\cdot \boldsymbol{m}_i/N_{\mathrm{side}}}\,,\\
\label{eq:delta_nr}
\end{split}
\end{equation}
which can be calculated by creating a new FFT box containing the data
$\Delta(\boldsymbol{k}_i)$ wherever a pixel vector meets the
requirement that $\boldsymbol{k}_i/k_{\mathrm{f}} \simeq
\boldsymbol{m}_i$, and zero otherwise. Then this new FFT box can
be FFTed to real space to create $\delta(\boldsymbol{n},\,\boldsymbol{k}_i)$.
Equivalently, a new FFT box can be created containing 1 wherever
$\boldsymbol{k}_i/k_{\mathrm{f}} \simeq \boldsymbol{m}_i$, and zero
otherwise, which may then be FFTed to real space to generate
$I(\boldsymbol{n},\,\boldsymbol{k}_i)$.
Our estimator for the polyspectrum can now be reduced to,
\begin{equation}
\begin{split}
&\mathscr{P}(\vect{k}_1, \vect{k}_2, ... \vect{k}_p) \approx
H^p\frac{1}{V}\frac{\sum\limits_{\boldsymbol{n}}^{N_{\mathrm{pix}}}
\prod \limits_{i=1}^p \delta(\boldsymbol{n},\,\boldsymbol{k}_i)}
{\sum\limits_{\boldsymbol{n}}^{N_{\mathrm{pix}}}\prod \limits_{i=1}^p I(\boldsymbol{n},\,\boldsymbol{k}_i)}\,,\\
\label{eq:Jeong_gen}
\end{split}
\end{equation}
The product within the summations is equivalent to performing
an inverse-FFT of a convolution in k-space as $\mathrm{FFT}[g(x)h(x)]
= g(k)*h(k)$.

Until this point we have described how FFTs may be used to implement
Equation \ref{eq:Jeong_gen}. As FFTs assume a real dataset,
a $\mathscr{P}(\vect{k}_1, \vect{k}_2, ... \vect{k}_p)$ resulting from using
FFTs will be a real quantity.
However, Equation \ref{eq:Jeong_gen} can equally be applied to complex datasets
by using complex DFTs (discrete FT) instead of FFTs.

The power spectrum may be calculated using the FFT-polyspectrum
estimator as follows,
\begin{equation}
\begin{split}
 P(\vect{k}_1, \vect{k}_2) &\approx
\frac{V}{N_{\mathrm{pix}}^2}
\frac{\sum\limits_{\boldsymbol{n}}^{N_{\mathrm{pix}}}
\delta(\boldsymbol{n},\,\vect{k}_1)\delta(\boldsymbol{n},\,\vect{k}_2)}
{\sum\limits_{\boldsymbol{n}}^{N_{\mathrm{pix}}}
I(\boldsymbol{n},\,\vect{k}_1)I(\boldsymbol{n},\,\vect{k}_2)}\,,\\
P(k_1) &\approx
\frac{V}{N_{\mathrm{pix}}^2}\frac{\sum\limits_{\boldsymbol{n}}^{N_{\mathrm{pix}}}
\delta(\boldsymbol{n},\,k_1)\delta(\boldsymbol{n},\,k_1)}
{\sum\limits_{\boldsymbol{n}}^{N_{\mathrm{pix}}}
I(\boldsymbol{n},\,k_1)I(\boldsymbol{n},\,k_1)}\,,\\
\label{eq:Jeong_PS}
\end{split}
\end{equation}
where in the second line we have made the standard assumption that
because the Universe is homogeneous and isotropic, the power spectrum
only depends on the separation of two points in real space, i.e. the
magnitude of a single $k$-mode. It is worth noting that in the case of
the spherically-averaged power spectrum it is actually faster to use
direct measurement rather than the FFT power-spectrum estimator as,
in this case, direct measurement only involves a single loop through the
box.
It is therefore only worth considering using the FFT
polyspectrum estimator when calculating higher-order statistics.

Equivalently, the bispectrum may be estimated by,
\begin{equation}
\begin{split}
& B(k_{\textsc{f}}\vect{m}_1, k_{\textsc{f}}\vect{m}_2, k_{\textsc{f}}\vect{m}_3)\\
& \approx \frac{V^2}{N_{\mathrm{pix}}^3}
\frac{\sum\limits_{\boldsymbol{n}}^{N_{\mathrm{pix}}}
\delta(\boldsymbol{n},\,\vect{k}_1)\delta(\boldsymbol{n},\,\vect{k}_2)\delta(\boldsymbol{n},\,\vect{k}_3)}
{\sum\limits_{\boldsymbol{n}}^{N_{\mathrm{pix}}}
I(\boldsymbol{n},\,\vect{k}_1)I(\boldsymbol{n},\,\vect{k}_2)I(\boldsymbol{n},\,\vect{k}_3)}\,,\\
\label{eq:Jeong_BI}
\end{split}
\end{equation}
In essence, we have reduced our bispectrum calculation from an
expensive nested loop though the FFT box, to one and a half loops
through the dataset (i.e. $3N_{\mathrm{pix}}/2$ pixels) and six (or for
a $p^{\mathrm{th}}$-order polyspectra, $2\,p$) FFTs, which are trivial
to parallelise with openMP.
The FFT-estimator's speed means that it is well suited to sampling applications.
Another useful feature of the FFT-estimator is that there is very little overhead to
calculating all the $p<\mathscr{P}$ spectrum, e.g. if you calculate
the trispectrum ($\mathscr{P}=4$), you can get the bispectrum ($p=3$)
and power spectrum ($p=2$) for the $k$-modes of the given trispectrum configuration
at no extra cost.

In implementing the FFT-estimator numerically, it is possible to
improve performance by making an initial pass through the whole box,
to build an indexing array in which the $j^\mathrm{th}$ entry
contains the dimensionless co-ordinates $m_x, m_y, m_z$ (cast to 1D)
of all pixels in the box for which $|j - A\,\sqrt{m_x^2 + m_y^2 +
  m_z^2}|<1/2$.
We introduce an integer scale factor $A$, without which the sampling is too coarse
and the performance of the estimator is impacted.
We set the scale factor
$A=1000$, and find this produces fine enough sampling to reproduce the
results produced by loading $\delta(\boldsymbol{n},\,\vect{k}_i)$ with
a full loop through the box each time.  Using the indexing array,
filling a given $\delta(\boldsymbol{n},\,\vect{k}_i)$ box only requires
loading the pixels whose co-ordinates are contained in the $j$ indexes satisfying
$|j - A\,(\sqrt{k_x^2 + k_y^2 + k_z^2}/k_{\mathrm{f}})|<s/2$.
Another point to note is that, as the method depends heavily on FFTs,
it notably maximizes the efficiency of the code to use a resolution
of $2^n$ on a side and to use threading with
openMP when executing FFT plans.
On a MacBook Pro with a Intel Core i5 (2.9 GHz) dual-core processor,
a single measurement of $B(\boldsymbol{k}_1,\, \boldsymbol{k}_2,\, \boldsymbol{k}_3)$
using the FFT-estimator bispectrum algorithm (and including the
indexing-array approach and openMP-threaded FFTs)
from a cubic box with 512 pixels per side takes about 10 seconds.

\section{The FFT bispectrum estimator - comparisons with the direct-measurement
method and theoretical predictions} \label{sec:Test}

\begin{figure}
  \centering
    \includegraphics[trim=0.0cm 0.0cm 0.0cm 0.0cm, clip=true, scale=0.55]{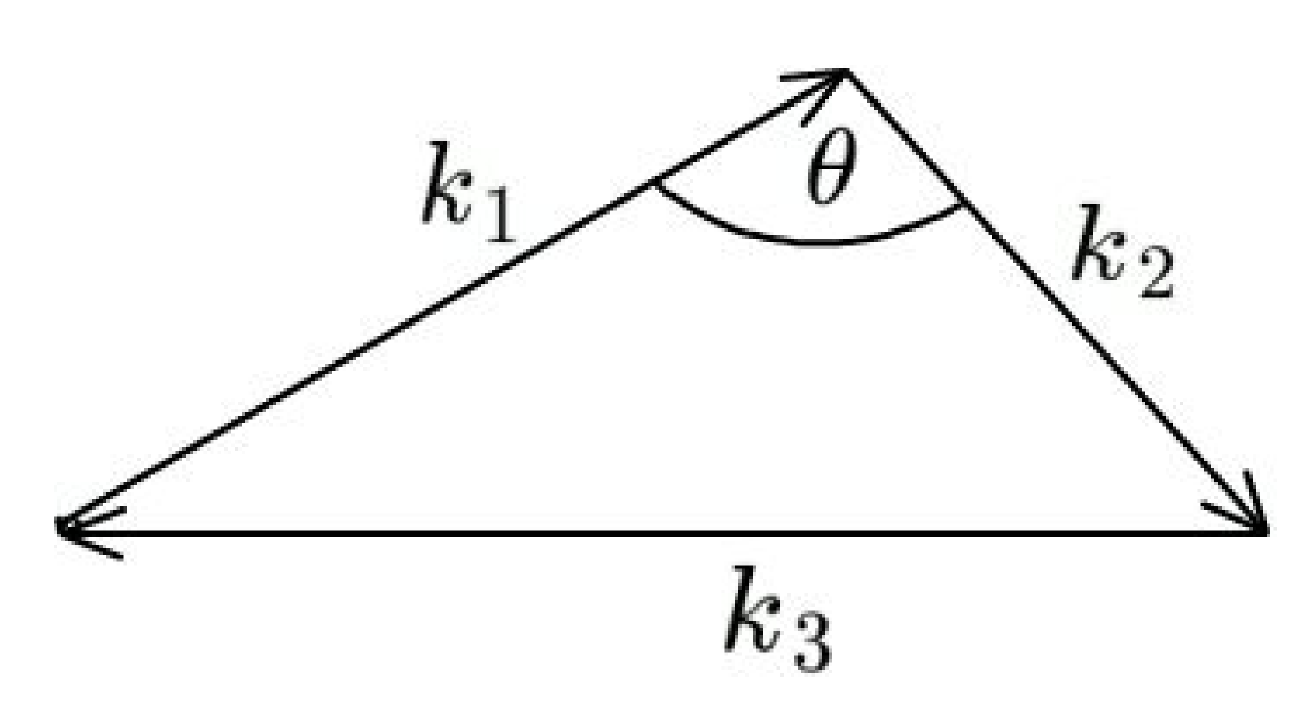}\\
  \caption{
  Illustration of the angle plotted throughout this paper, with
  respect to the vectors $\vect{k}_1$ and $\vect{k}_2$, where
  $\vect{k}_3 = - \vect{k}_1 - \vect{k}_2$ closes the triangle.}
  \label{fig:theta_illus}
\end{figure}

To better understand, and to test, the FFT algorithm we present in this
work, we compare the FFT estimator, as applied to the power spectrum and bispectrum,
with a direct-measurement method.\footnote{It is worth noting,
that \citet{Sefusatti2015} compare measuring the
bispectrum from the Fourier modes of an N-body simulation (which do not
suffer from aliasing), with that measured by first gridding the particles,
applying an FFT, and then applying the FFT estimator.
This determines the impact of aliasing, but does not compare direct
and FFT bispectrum measurements from gridded datasets.}
For our tests we choose a slightly non-Gaussian dataset, namely a non-linearly evolved
density field, and a very non-Gaussian dataset in the form of a toy
model for reionization. In the raw measurements of the FFT bispectrum, we
use a bin width of $s=3$, because throughout we measure the
spherically-averaged bispectrum and $s=3$ accounts for modes within a
pixel distance of the components constructing a given
$|\vect{k}|/k_{\mathrm{f}}$, i.e. $s/2 \sim \sqrt{3(1^2)}$.\footnote{Note
that we find that using a fixed bin width works better than all the
variable bin widths we considered.
We consider $\d k = s\,k_f\,k/2$ but this works very badly as the bins
are too big at large-$k$ and too small at small-$k$. Worse still is
$\d k = s\,k_f/(2\,k)$.
In general the chosen bin
width will cause the estimator to breakdown below a certain $k$;
for example, choosing $s=4$ would mean that the FFT estimator will break down
for $k/k_f/<s\pi/L$ = 0.02 when $L=600$ Mpc.}
In many of the plots we present in this paper we plot the bispectrum as a
function of $\theta$, which corresponds to the internal angle between
vectors $\vect{k}_1$ and $\vect{k}_2$ when they are added,
this is illustrated in Figure \ref{fig:theta_illus}.

Throughout the paper we compare the FFT-bispectrum measurements to
that of theory, but also to the bispectrum from a direct-measurement method.

\subsection{Direct measurement of the bispectrum}
\label{sec:direct_est}
To evaluate the performance of the FFT bispectrum estimator, it is desirable to
draw comparison with another algorithm.
We therefore use a restricted implementation of the direct-measurement method,
which has been designed to reduce calculation time, and make the measurements
presented here computationally tractable.

The main reason one would like to have a faster estimator for
the bispectrum, or any other higher order polyspectra,
is because the conventional direct estimators
(that directly implement Equation \ref{eq:bi_definition} in their
algorithm) of such polyspectra, require a
significant amount of computational time.
To implement Equation \ref{eq:bi_definition} in the direct
algorithm of bispectrum, one would typically need to go through
six nested for loops,{\footnote{To
    construct all possible vector triplets ($\vect{k}_1,\,\vect{k}_2,\,\vect{k}_3$
    in a three dimensional vector space) in the FT
    box, one would need nine nested for loops.  However, when we
    impose the condition that these vector triplets should form a
    closed triangle, that reduces it to six nested for loops. The
    equation of constraint ($\vect{k}_1+\vect{k}_2+\vect{k}_3 = 0$) in
    this case is a vector equation, thus effectively three scalar
    equations and reduces three degrees of freedom.}}
each the size  of the FFT box side in grid
units.{\footnote{If the actual
  field, $V(\bf x)$, for which one wants to estimate the
  polyspectra is real, due to its Hermitian properties, only half of
  the Fourier space will contain unique information about the field
  and the other half can be created using the condition
  $V^{*}({\bf k}) = V(-{\bf k})$.}}
  Such a nested loop is very computationally expensive.

To reduce the number of nested loops,
we introduce two constraints on $\vect{k}_1$ and $\vect{k}_2$ in our direct-estimation algorithm.
For a specific kind of
triangle configuration, the ratio between the two arms of the triangle
must remain constant, i.e.
\begin{equation}
\label{eq:ratio}
k_2/k_1 = m    \,,
\end{equation}
and the cosine of the angle ($\alpha = \pi - \theta$) between the two
vector arms of the triangle must be fixed to,
\begin{equation}
\label{eq:alpha}
\frac{\vect{k}_1\cdot\vect{k}_2} {k_1 k_2} = \cos\alpha \,.
\end{equation}
Implementation of these two 
constraints in the algorithm
requires four nested for loops rather than six.
This reduces the total number of steps in the algorithm to $N^4$, instead of $N^6$.
where $N$ is the number of steps corresponding to each for loop.

In this algorithm,
the first three for loops determine all possible values of
the three components of the $\vect{k}_1$ vector, and the fourth for loop
determines all possible values of the one component of the $\vect{k}_2$
vector.
The other two components of the $\vect{k}_2$ vector are
fixed by Equations \ref{eq:ratio} and \ref{eq:alpha} for a given
$\vect{k}_1$ vector, and a single component of the $\vect{k}_2$ vector.
The $\vect{k}_3$ vector is determined using the closure
condition of the triangle.
Once all components of $\vect{k}_1$, $\vect{k}_2$
and $\vect{k}_3$ vectors are determined, one can take the
product of the $\Delta(\vect{k})$s corresponding to these three
vectors, which will be a complex number (as are all
$\Delta({\bf k})$s).
If the actual field for which
one intends to estimate the bispectrum is real it can easily be shown
(using complex algebra and the Hermitian condition mentioned before)
that the bispectrum will also be real.
Thus, we take only the real part of this complex product as
our bispectrum contribution to each bin.
We also estimate the power spectrum contribution
from each of the three arms of the triangle in three separate bins,
corresponding to $P(k_1)$, $P(k_2)$ and $P(k_3)$. In these power
spectrum bins only $k$ vectors which satisfy the
closure condition of Equation \ref{eq:bi_definition} contribute,
and we use these $P(k)$s to estimate the Perturbation theory
expectation for the bispectrum of N-body density fields as described
by Equation \ref{eq:PT_Bi} in Section \ref{PT}.

This particular algorithm for direct estimation of bispectrum is very
restrictive in nature when compared to the fast algorithm upon which
this paper is focused.
While the fast algorithm allows any kind of bin width around the target $\vect{k}_1$,
$\vect{k}_2$ and $\vect{k}_3$ vectors, corresponding to a specific
triangle configuration, in this direct algorithm one can only put a
bin width around $\vect{k}_1$ but it is not possible to put any bin
widths around $\vect{k}_2$ and $\vect{k}_3$, as their components are
determined precisely by Equations \ref{eq:ratio}, \ref{eq:alpha} and
the closure condition of a triangle for a specific set of components
of $\vect{k}_1$. Due to this difference in the nature of binning
in these two algorithms,
they will be probing bispectrum for a slightly different
sets of triangles, when averaged across their respective $k$ bins. We
thus do not expect a direct one-to-one exact match/correspondence
between these two methods while comparing the bispectrum estimated by
them.

\subsection{Non-linear density field - A slightly non-Gaussian test case}
\label{PT}

In testing our FFT estimator, it is useful to have theoretical predictions of
the bispectrum with which to draw comparison.
As such, it is useful to consider the bispectrum of the density field.

Perturbation theory describes the initial density field with a
background term, and perturbative terms. Whilst the background term
will have a vanishing three-point correlation function and bispectrum,
the perturbative terms which evolve in a non-linear manner under
gravity will exhibit non Gaussianities. \citet{Fry1984} use
perturbation theory, to second order (or tree level), to make a prediction for the
$\vect{k}$ dependence on the bispectrum of the matter density field,
finding that,
\begin{equation}
\begin{split}
B(k_1, k_2, k_3) &= 2F(k_1,k_2)P(k_1)P(k_2) + (\mathrm{cyc.})\\
F(k_1,k_2) &= \left(\frac{1+\kappa}{2}\right)
+ \left( \frac{\vect{k}_1\cdot\vect{k}_2}{2k_1k_2}\right)
\left(\frac{k_1}{k_2} + \frac{k_2}{k_1}\right)\\
&+ \left( \frac{1-\kappa}{2}\right)
\left( \frac{\vect{k}_1\cdot\vect{k}_2}{k_1k_2}\right)^2
\,,\\
\label{eq:PT_Bi}
\end{split}
\end{equation}
where $\kappa = 3/7\,\Omega_{\mathrm{m}}^{-1/143}$ as appropriate for
a $\Lambda$CDM cosmology \citep{Scoccimarro2000}.
This tree-level bispectrum prediction has been shown to under predict the
bispectrum as measured from N-body simulations.
This is especially true for scales corresponding to strongly non-linear scales,
but theory still under predicts the N-body bispectrum on scales for which density
fluctuations are small and still non-linear,
e.g. \citep{Scoccimarro1997}.
To compare our FFT-estimator measurements of the bispectrum with the
predictions of tree-level perturbation theory,
we use the Particle-Mesh N-body matter density simulations described
by \citet{Mondal2014} and \citet{Bharadwaj2004}.
This simulation was run with a $4288^3$ grid, and a cube side of 300 Mpc.
This provides a spatial resolution of $\sim 0.07$ Mpc, and mass resolution
$1.09\times 10^8$ M$_\odot$.
The boxes we analyse here have been coarse gridded to $536^3$.

\begin{figure}
  \centering
  $\renewcommand{\arraystretch}{-0.75}
  \begin{array}{c}
    \includegraphics[trim=1.5cm 2.65cm 0.5cm 2.1cm, clip=true, scale=0.25]{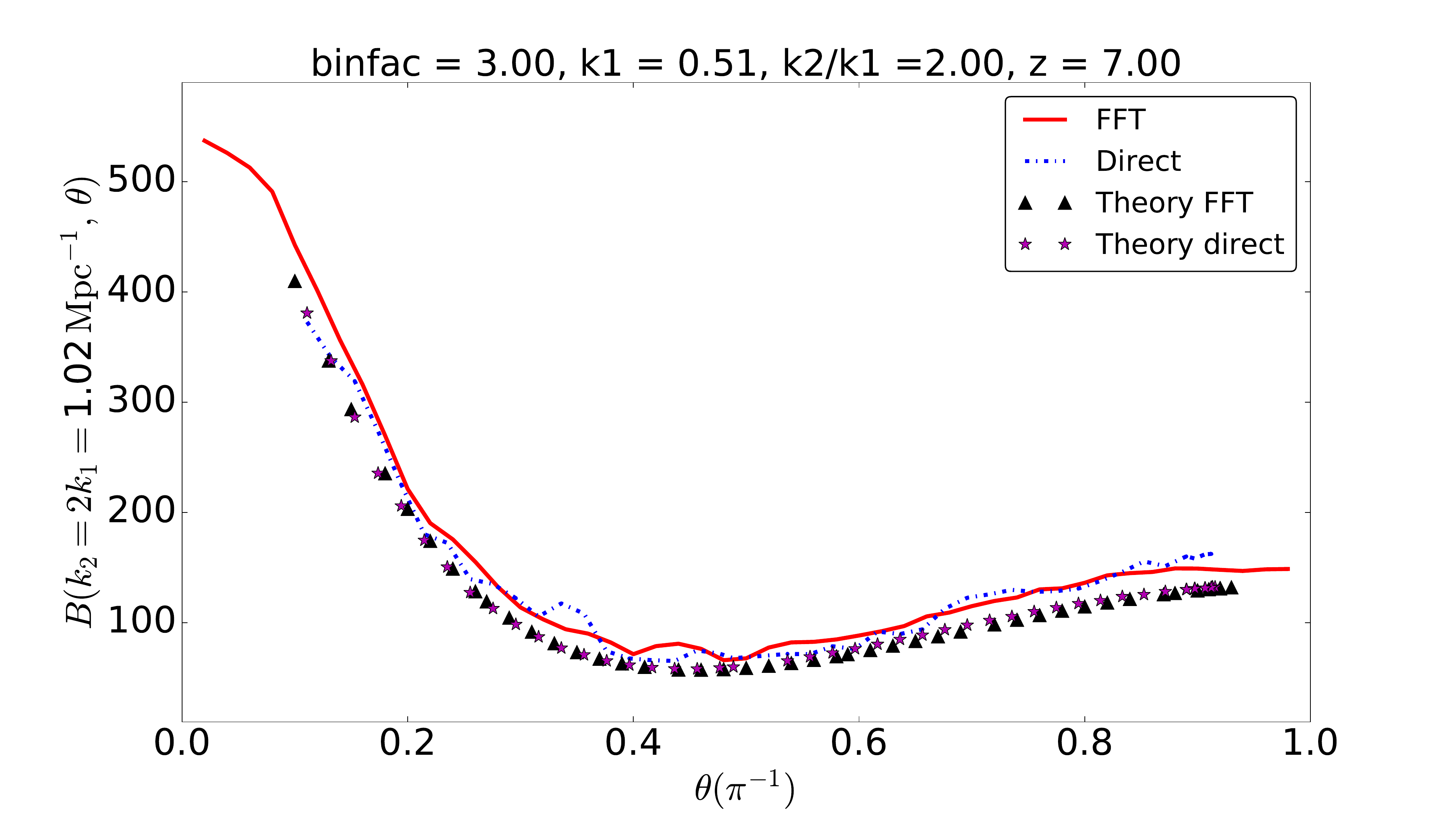}\\
    \includegraphics[trim=1.5cm 2.65cm 0.5cm 2.1cm, clip=true, scale=0.25]{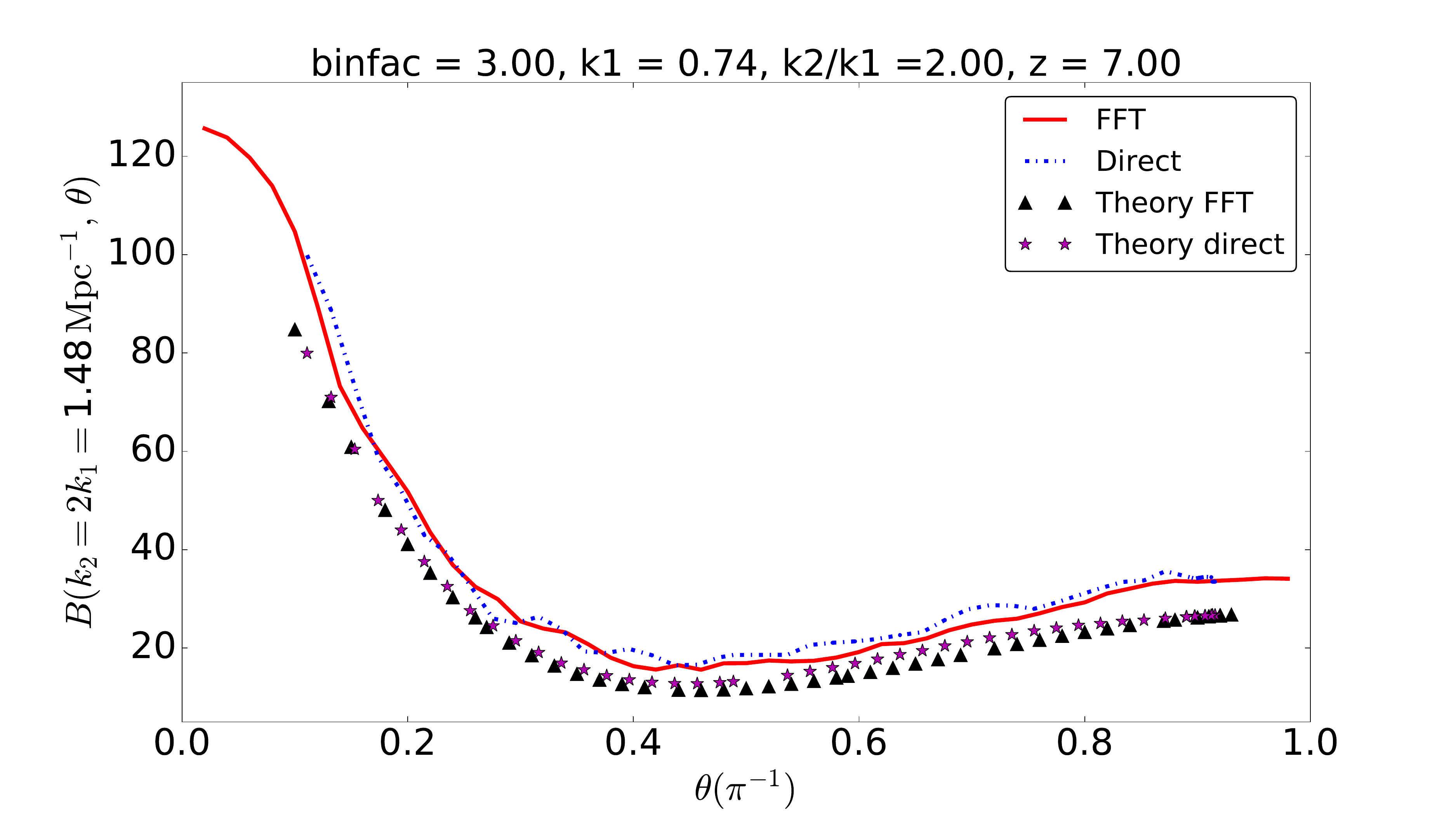}\\
    \includegraphics[trim=1.5cm 0.5cm 0.5cm 2.1cm, clip=true, scale=0.25]{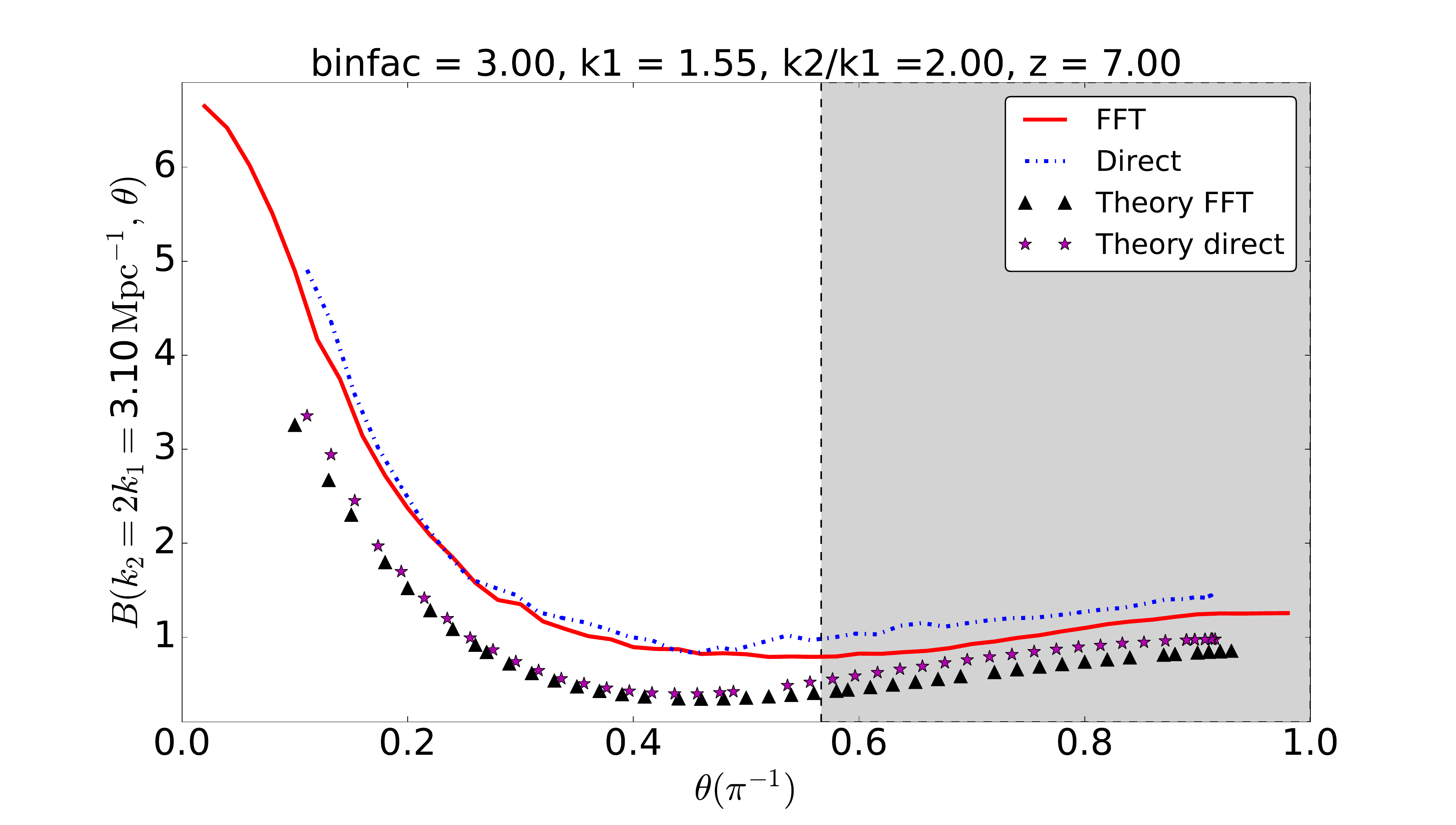}\\
  \end{array}$
  \caption{Bispectrum measured from a non-linearly evolved density field.
  Pink stars mark the theoretical prediction as calculated using the same
  binning as the direct method.
  We plot the bispectrum as a function of angle between $k_1$ and $k_2$
  where  $k_2= 2\,k_1$.
  From top to bottom we plot $k_1 = (0.51, 0.74, 1.55)\,\mathrm{Mpc}^{-1}$
  for which $k_2 = (1.02, 1.48, 3.10)\,\mathrm{Mpc}^{-1}$ respectively.
  The grey shaded area corresponds to $k$ values beyond which \citet{Sefusatti2015}
  predict that the FFT bispectrum estimator will become inaccurate.
  Beyond divergence due to binning differences (clear by comparing the PT
  predictions under the two different binning schemes), the FFT estimator
  performs well, even in the grey shaded region.
  }
  \label{fig:density_k2eq2k1}
\end{figure}

\begin{figure}
  \centering
  $\renewcommand{\arraystretch}{-0.75}
  \begin{array}{c}
    \includegraphics[trim=1.75cm 2.65cm 0.5cm 2.1cm, clip=true, scale=0.25]{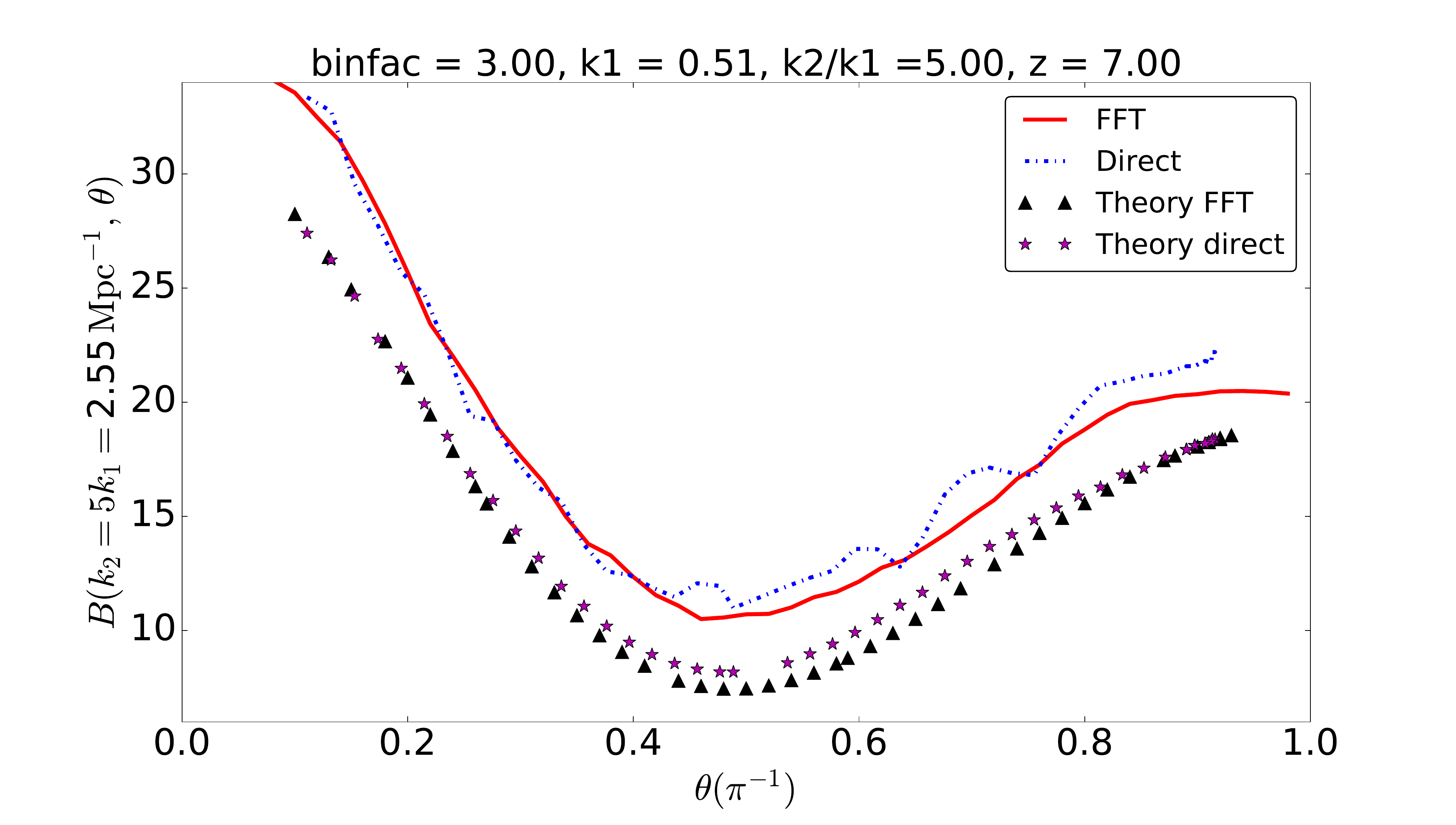}\\
    \includegraphics[trim=1.75cm 0.5cm 0.5cm 2.1cm, clip=true, scale=0.25]{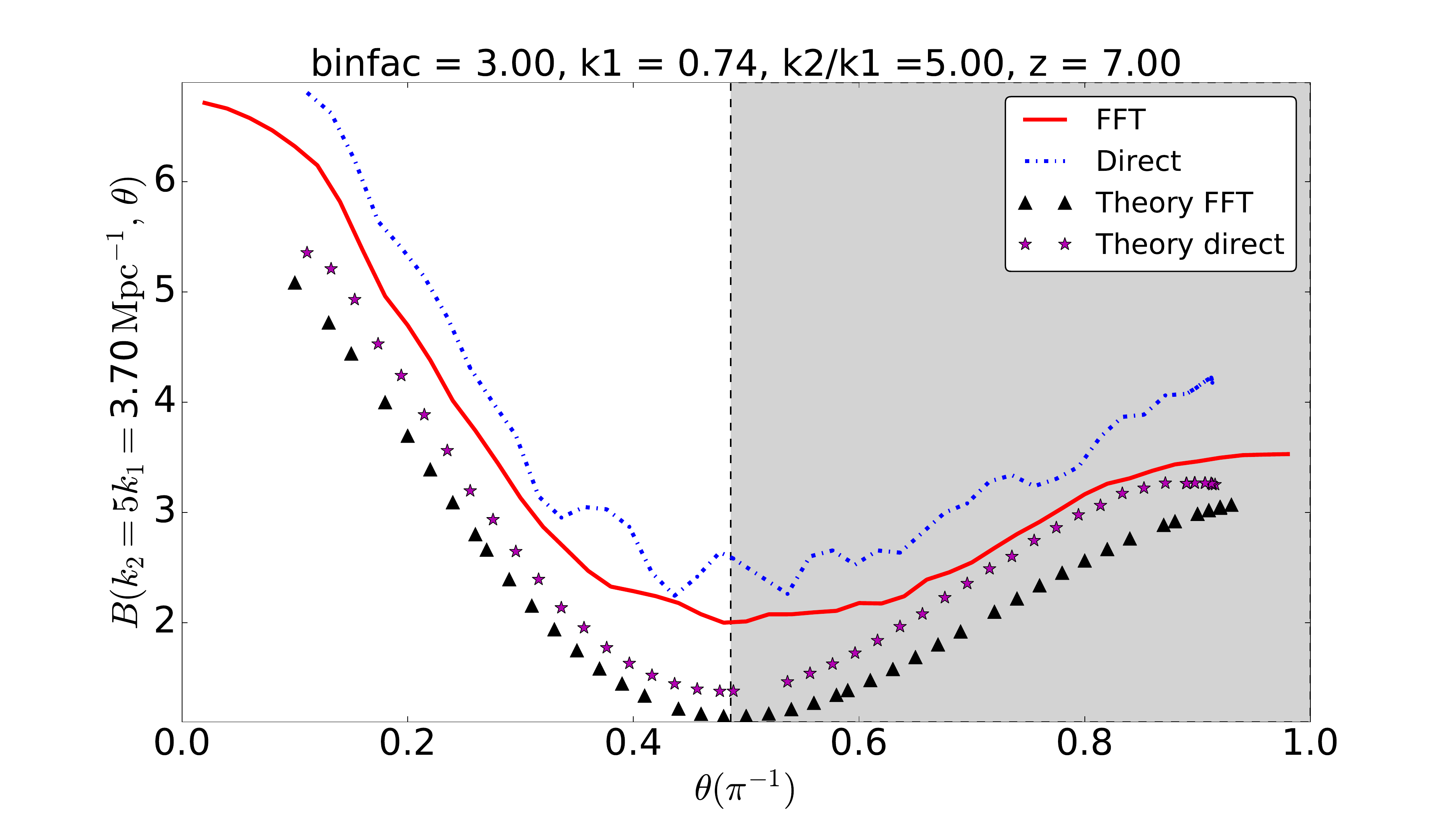}\\
  \end{array}$
  \caption{
  Bispectrum measured from a non-linearly evolved density field.
  Pink stars mark the theoretical prediction as calculated using the same
  binning as the direct method.
  We plot the bispectrum as a function of angle between $k_1$ and $k_2$
  where  $k_2= 5\,k_1$.
  From top to bottom we plot $k_1 = (0.51, 0.74)\,\mathrm{Mpc}^{-1}, k_2 = (2.55, 3.70)\,\mathrm{Mpc}^{-1}$ respectively,
  note that we cannot plot $k_1 = 1.55\,\mathrm{Mpc}^{-1}$ as this pushes $k_2$ beyond the
  Nyquist limit.
  The grey shaded area corresponds to $k$ values beyond which \citet{Sefusatti2015}
  predict that the FFT bispectrum estimator will become inaccurate.
  Again, the FFT estimator is seen to perform very well as compared to theory
  and our direct method.
  }
  \label{fig:density_k2eq5k1}
\end{figure}

\begin{figure}
  \centering
  $\renewcommand{\arraystretch}{-0.75}
  \begin{array}{c}
    \includegraphics[trim=1.25cm 2.67cm 0.5cm 2.1cm, clip=true, scale=0.25]{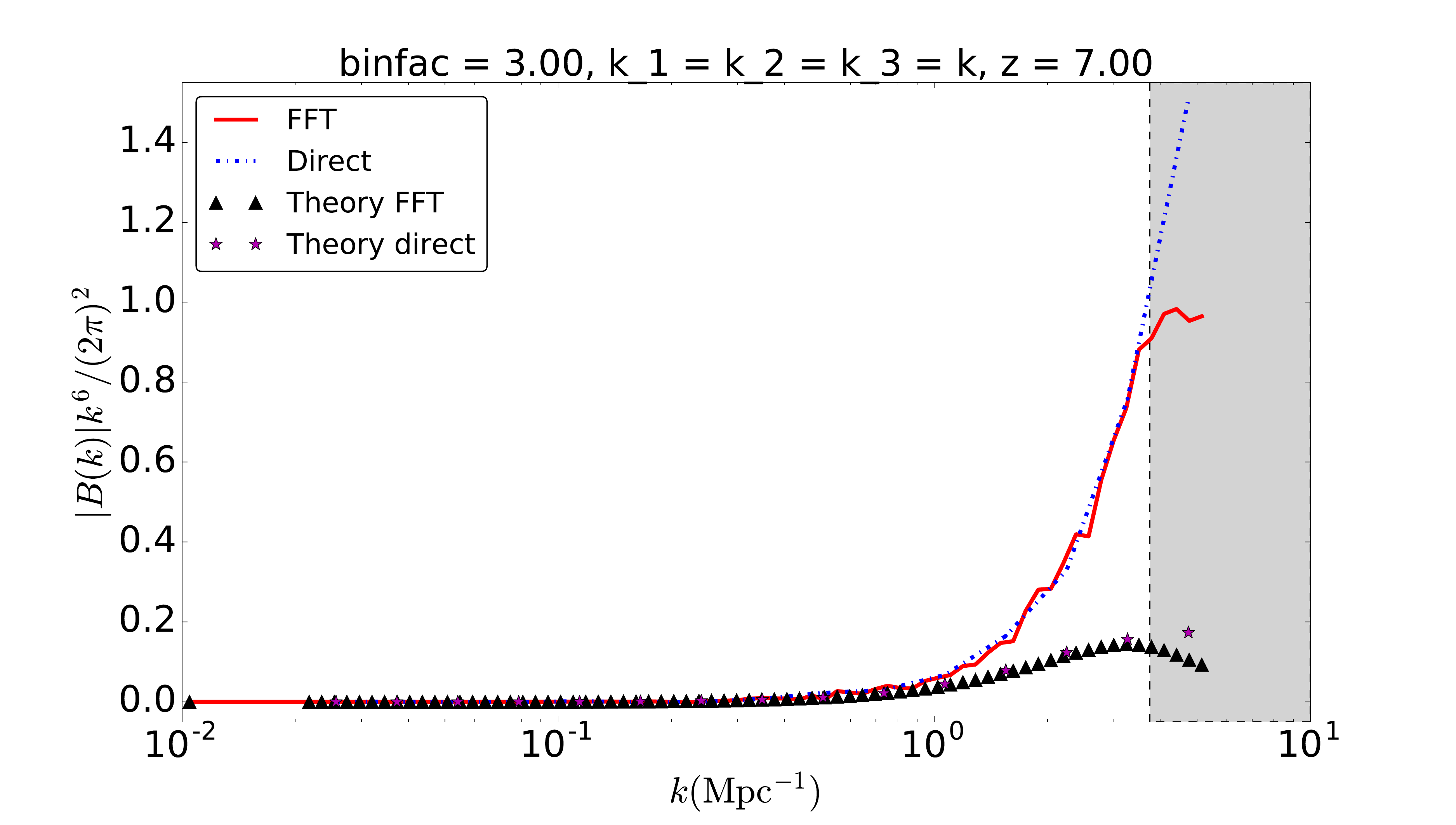}\\
    \includegraphics[trim=1.25cm 0.3cm 0.5cm 2.1cm, clip=true, scale=0.25]{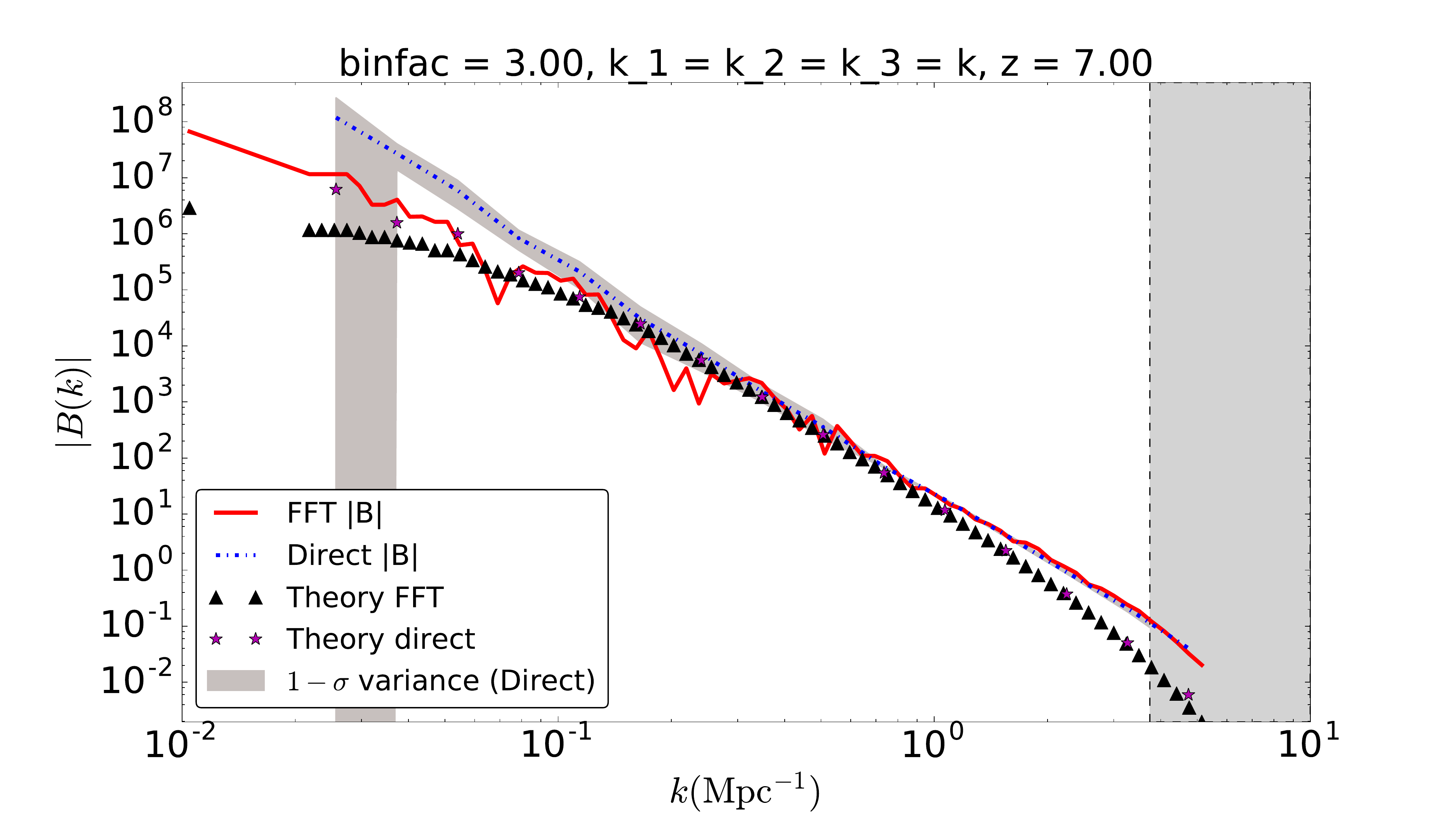}\\
  \end{array}$
  \caption{
  Bispectrum measured from a non-linearly evolved density field.
  Pink stars mark the theoretical prediction as calculated using the same
  binning as the direct method.
  We plot the bispectrum as a function of $k_1$ for triangles where $k_1=k_2=k_3$.
  We show $B(k_1=k_2=k_3)\,k_1^6/(2\pi^2)^2$ in the top plot.
  In the bottom plot, we plot the average $|B(k)|$ across
  5 different realisations as measured by the direct method
  (blue dot-dashed line).
  The beige shaded region marks the 1-$\sigma$ standard deviation across the
  5 realisations.
  For this simulation, we find that $B(k)$ exhibits erratic evolution
  due fluctuations in sign caused by sample variance,
  the amplitude of the real part of the bispectrum is far more stable.
  The grey shaded area corresponds to $k$ values beyond which \citet{Sefusatti2015}
  predict that the FFT bispectrum estimator will become inaccurate.
  Other than differences from noise due to sample variance and differences in binning,
  the FFT estimator is seen to perform very well as compared to theory and our direct method.
  }
  \label{fig:density_bi_equi}
\end{figure}

In Figures \ref{fig:density_k2eq2k1}
to Figure \ref{fig:density_bi_equi} we plot the bispectrum, from a density
simulation at $z=7$, as measured using the FFT bispectrum estimator of
Equation \ref{eq:Jeong_BI} (red solid line), the direct-measurement method
described at the beginning of this
section (blue dot-dashed line) and as predicted by PT (black triangles), i.e. Equation
\ref{eq:PT_Bi}.
To highlight divergence between the direct and FFT methods due to differences their $k$ binning,
we also plot the PT prediction binned as per our direct-measurement
method (pink stars).
We plot the bispectrum as a function of angle ($\theta$ in $\pi^{-1}$ radians)
for $k_2 = 2\,k_1$ in Figure \ref{fig:density_k2eq2k1} and for $k_2 =
5\,k_1$ in Figure \ref{fig:density_k2eq5k1}, with $k_1 = (0.51, 0.74,
1.55)\,\mathrm{Mpc}^{-1}$ from top to bottom (note that in Figure \ref{fig:density_k2eq5k1} we do
not plot $k_1 = 1.55\,\mathrm{Mpc}^{-1}$ as $k_2$ is greater than the Nyquist limit).
Here we average over bins of $\cos(\theta)\pm 0.05$ for both direct and
FFT estimators.
\footnote{We choose to
bin in $\cos{\theta}$ as our direct method samples $\cos({\pi - \theta})$ in
linear bins.}

From these figures it is clear that the FFT estimator closely follows
the PT theoretical predictions, only diverging on smaller scales
(larger $k$-modes) as expected.  The direct-measurement method also
agrees well with the FFT estimator.
We note that there is some divergence between the two methods for
$k_2 = 3.10\,\mathrm{Mpc}^{-1}$ for $\theta \gtrsim 0.5\, \pi^{-1}$ radians,
which is due to differences in the binning between the two methods.
This is clear as we see the same qualitative divergence
between the theoretical predictions resulting from each method's binning.

The Nyquist theorem states that the smallest wavelength that may be resolved
is 2 samples (in our case pixels),
this corresponds to a limit on $k$ of,
\begin{equation}
\begin{split}
k_{\mathrm{nyq}} = \frac{2\pi}{l_{\mathrm{nyq}}}
= \frac{2\pi}{(2L/N_{\mathrm{side}})}
= k_{\textsc{f}}N_{\mathrm{side}}/2\,.
\label{eq:nyqvist}
\end{split}
\end{equation}
We therefore do not calculate the bispectrum for triangles that incorporate
any $|\vect{k}|\ge k_{\mathrm{nyq}}$.
However, \citet{Jeong2010a} conclude the largest mode for which the FFT bispectrum estimator
is stable (i.e. not affected by aliasing) is three times smaller than the 1D FFT grid, or
$k = N_{\mathrm{side}}\,k_{\mathrm{f}}/3$.
We mark this limit on all plots by a grey shaded region.
This conclusion is reached by counting the triangles using the
FFT approach (i.e. applying the denominator of Equation \ref{eq:Jeong_BI}),
and comparing it to the true counted value.

An alternative theoretical argument for this limit,
that relates to aliasing, is provided in \citet{Sefusatti2015}.
In equation \ref{eq:delta_nr}, we are essentially performing the following
operation,
\begin{equation}
\begin{split}
&B(\boldsymbol{k}_{1}, \boldsymbol{k}_{2}, \boldsymbol{k}_{3}) =\\
&\frac{1}{N_{\mathrm{tri}}}\int d^3\vect{x}\int_{k_1} d^3\vect{q_1}\int_{k_2} d^3\vect{q_2}\int_{k_3} d^3\vect{q_3}\delta_{\boldsymbol{q}_1}\delta_{\boldsymbol{q}_2}\delta_{\boldsymbol{q}_3}
\mathrm{e}^{i\,\boldsymbol{q}_{123}\cdot\boldsymbol{x}}\,,\\
\label{eq:delta_sefusatti}
\end{split}
\end{equation}
where $\boldsymbol{q}_{123} =  \boldsymbol{q}_{1} + \boldsymbol{q}_{2} + \boldsymbol{q}_{3}$,
and the integrals are over grid points for which $\boldsymbol{q}_{i} = \boldsymbol{k}_{i}\pm \Delta k$
(where $\Delta k$ is the chosen bin width).
\citet{Sefusatti2015} argue that the exponent in this expression is invariant under a 1-dimensional
translation of each wavenumber of $(2\pi/L)\,(N_{\mathrm{side}}/3)$ for which
$\boldsymbol{q}_{123}
\rightarrow \boldsymbol{q}_{123} + 2\pi N_{\mathrm{side}}/L$.
The translation cancels with $\boldsymbol{x} = (L/N_{\mathrm{side}}) \boldsymbol{m}$,
introducing a factor of $\mathrm{exp}(i\, 2\pi \boldsymbol{m})$.
As $\boldsymbol{m}$ is an integer triplet the exponent associated with the transpose
is always one.
The argument is that this means that there is a periodicity in the phase term
associated with this translation scale, which defines a maximum wavenumber,
$k_{\mathrm{max}} = N_{\mathrm{side}}\,k_{\mathrm{f}}/3$, beyond which the
estimator will become confused.
If this argument stands, then $k_{\mathrm{max}}$ will decrease
according to $k_{\mathrm{max}}(p) = N_{\mathrm{side}}\,k_{\mathrm{f}}/p$ for
a $p^{\mathrm{th}}$-order polynomial.

In the results that follow,
it appears that this confusion effect does not
seem to seriously affect the performance of the estimator,
at least for the datasets considered here.
If we were to exactly implement a Dirac-delta function using a Fourier transform,
as per Equation \ref{eq:delta_sefusatti},
$\boldsymbol{q}_{123}\equiv 0$, which makes sure that the triangle is
closed, and so the exponential contribution is always 1.
In using the FFT estimator on a discrete dataset, this is not the case, as
$\boldsymbol{q}_{123}$ does not necessarily form a closed triangle, and so there
is `noise' introduced by the Kronecker-delta's exponential contribution not being
unity.
Any confusion due to the periodicity of the exponential phase term described above
(and originating from the FFT implementation of the Kronecker delta) must necessarily
be within the level of the `noise' inherent to the method as a whole.
To minimise noise introduced by the FFT implementation of the Kronecker delta,
we advocate using a bin width corresponding to one pixel when measuring the
bispectrum with the FFT estimator and, if required,
applying further binning subsequently.

When we consider the bispectrum normalised by $k^6/(2\pi^2)^2$
for the equilateral configuration, as shown in the top plot of Figure \ref{fig:density_bi_equi},
we see the bispectrum as measured by both direct and FFT methods,
diverges from the theoretical at $k \gtrsim 1$ Mpc$^{-1}$
(note that, for the direct method, we average over bins of
$\cos{\theta} = -0.5 \pm 0.05$).\footnote{We choose to
bin in $\cos{\theta}$ as our direct method samples $\cos({\pi - \theta})$ in
linear bins.}
This is not surprising as second-order perturbation theory cannot fully
describe the non-linearities of an N-body density field.
However, we also see that the FFT estimator and the direct-measurement method start
to diverge from each other at $k$ slightly lower than $k_{\mathrm{max}}$
(which is marked by the dashed line).
As the divergence does not start at exactly $k_{\mathrm{max}}$
and the theoretical predictions from the two methods also diverge in a qualitatively
very similar way. We conclude that it is, at least in part, due to differences in
binning between the two methods.
We also find that the impact of confusion due to periodicity of the phase term
of Equation \ref{eq:delta_sefusatti}
seems to be negligible in the case when two of the vectors that make up the triangle are
below $k_{\mathrm{max}}$.
This is clear from the bottom plots of Figures \ref{fig:density_k2eq2k1}
and \ref{fig:density_k2eq5k1} where we see the FFT-estimator and direct-measurement
method remain in reasonable agreement even for angles corresponding to
$k_3 \ge k_{\mathrm{max}}$.
There is slight divergence between the two methods, but it is more likely that this is
due to differences in binning, as, again, the same qualitative divergence is seen
when the theoretical predictions are binned as per each of the different methods.

At the other extreme of small-$k$ (large scales),
there is also a limit below which the triangle count
becomes too low, and the bispectrum gets impacted by sample variance.
We find this to occur when $N_{\mathrm{tri}}<10^7$, as measured using the FFT approach.
This corresponds to $k \lesssim (100/6)\,k_{\mathrm{f}}$,
i.e. when the $k_{\mathrm{f}}$ corresponds to greater than 6\% of the $k$ mode
under consideration.
Below this $k$ the estimators become increasingly noisy,
and the sign of the bispectrum also fluctuates from positive to negative at random.
This makes it very hard to interpret the signal, and where such wild fluctuations
are seen, we argue it is better to plot the absolute value of the bispectrum.
In Figure \ref{fig:density_bi_equi}, we plot the average of $|B(k)|$ as measured
by the direct method and its 1-$\sigma$ standard deviation (beige shaded region)
across 5 different realisations of the density field.
We find that that the impact of sample variance on $|B(k)|$ is less
dramatic than it is for $B(k)$; for illustration,
$\sim$50\% of $B[k<(100/6)\,k_{\mathrm{f}}]$
(from direct-measurement) of a single realisation have negative sign.{\footnote{
  The imaginary part of a bispectrum measured from a real field should be zero.
  However, this is not the case for the direct-measurement method as we only measure
  the bispectrum from half of $k$-space, which means that the imaginary contribution
  does not get cancelled out, as it would if we were to measure the bispectrum from the whole of $k$-space.
  Therefore in calculating $|B(k)|$ we take the absolute value of the real
  part.}}
Apparent from Figure \ref{fig:density_bi_equi} is a divergence between the two
methods at small $k$,
but again this may be attributed to differences in binning between the two methods.

\subsection{Toy-model for reionization - A highly non-Gaussian test case}
\label{sec:toy_model}
\begin{figure}
  \centering
  $\renewcommand{\arraystretch}{-0.75}
  \begin{array}{c}
    \fbox{\includegraphics[trim=0.0cm 0.0cm 0.0cm 0.0cm, clip=true, scale=0.19]{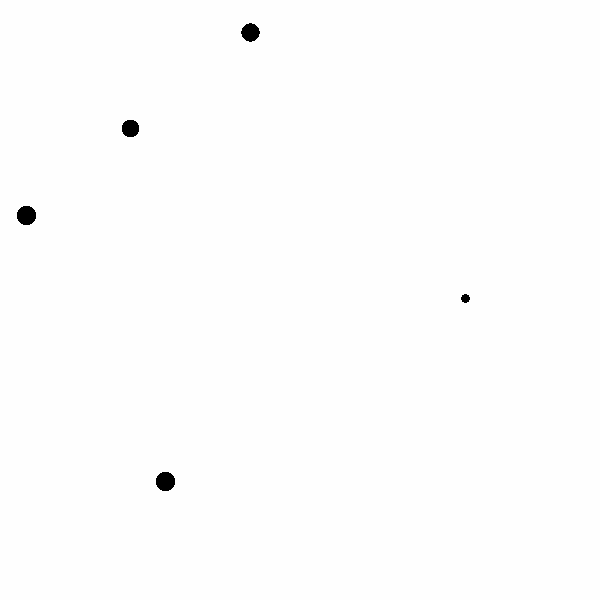}}\fbox{\includegraphics[trim=0.0cm 0.0cm 0.0cm 0.0cm, clip=true, scale=0.19]{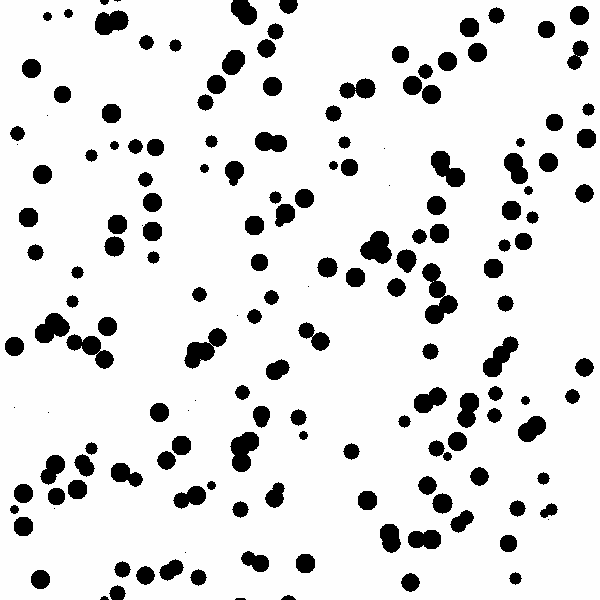}}\\
  \end{array}$
  \caption{Slices through the randomly-placed ionized spheres model for reionization;
  white depicts 100\% neutral regions and black 100\% ionized regions.
  The left column shows the models at $z=11$, $\xh = 0.99$ and the right
  column at  $z=14$, $\xh = 0.88$.
  For both the radius of all ionized spheres is 10 Mpc.
  }
  \label{fig:randomModel}
\end{figure}

\citet{Bharadwaj2005} present an analytical model for the bispectrum of the
ionization field during reionization.
To do so they assume that the ionized bubbles are randomly distributed
spheres, all of a single radius $R$ (where $R$ is a free parameter).
This radius is then used to define the number density of bubbles
$\overline{n}_{\hi}$ through $1-\overline{x}_{\hi} = (4\pi R^3/3)\overline{n}_{\hi}$,
with the neutral fraction $\overline{x}_{\hi}$ calculated according to the model of
\citet{Zaldarriaga2004}.\footnote{Note that the expression for $\overline{x}_{\hi}$ quoted
by \citet{Bharadwaj2005} is actually the expression for the ionized fraction.}
In this model the power spectrum of the ionization field is given by,

\begin{equation}
\begin{split}
P_{\hi}(k) =
\frac{(1-\overline{x}_{\hi})^2 W^2(kR)}{\overline{n}_{\hi}}\,,\label{eqn:BharPk}
\end{split}
\end{equation}
and the bispectrum by,

\begin{equation}
\begin{split}
B_{\hi}(k_1,k_2,k_3) =
-\frac{(1-\overline{x}_{\hi})^3 W(k_1R) W(k_2R) W(k_3R)}{\overline{n}_{\hi}^2}\,,\label{eqn:BharBk}
\end{split}
\end{equation}
where the window function $W(kR)$ is the Fourier transform of
the spherical top hat function.
We generate cubes that simulate the model of \citet{Bharadwaj2005}, so that we
may compare our estimator with the above theoretical predictions.
Slices through two simulation cubes are shown in
Figure \ref{fig:randomModel}. The left slice is at $z=14$ where reionization is
just beginning when $\xh = 0.99$;
the right figure is at $z=11$ when $\xh = 0.88$.
As we see from the right slice of Figure \ref{fig:randomModel},
the bubbles are in some cases overlapping with each other.
Such overlap is not allowed for in the model of \citet{Bharadwaj2005},
therefore we do not expect that the bispectrum measured from these boxes will exactly
agree with the theoretical predictions of Equations \ref{eqn:BharPk} and \ref{eqn:BharBk}.

\setlength\fboxsep{0pt}
\setlength\fboxrule{0.5pt}

\begin{figure}
  \centering
  $\renewcommand{\arraystretch}{-0.75}
  \begin{array}{c}
    \includegraphics[trim = 1.25cm 0.5cm 0.5cm 2.1cm, clip=true, scale=0.25]{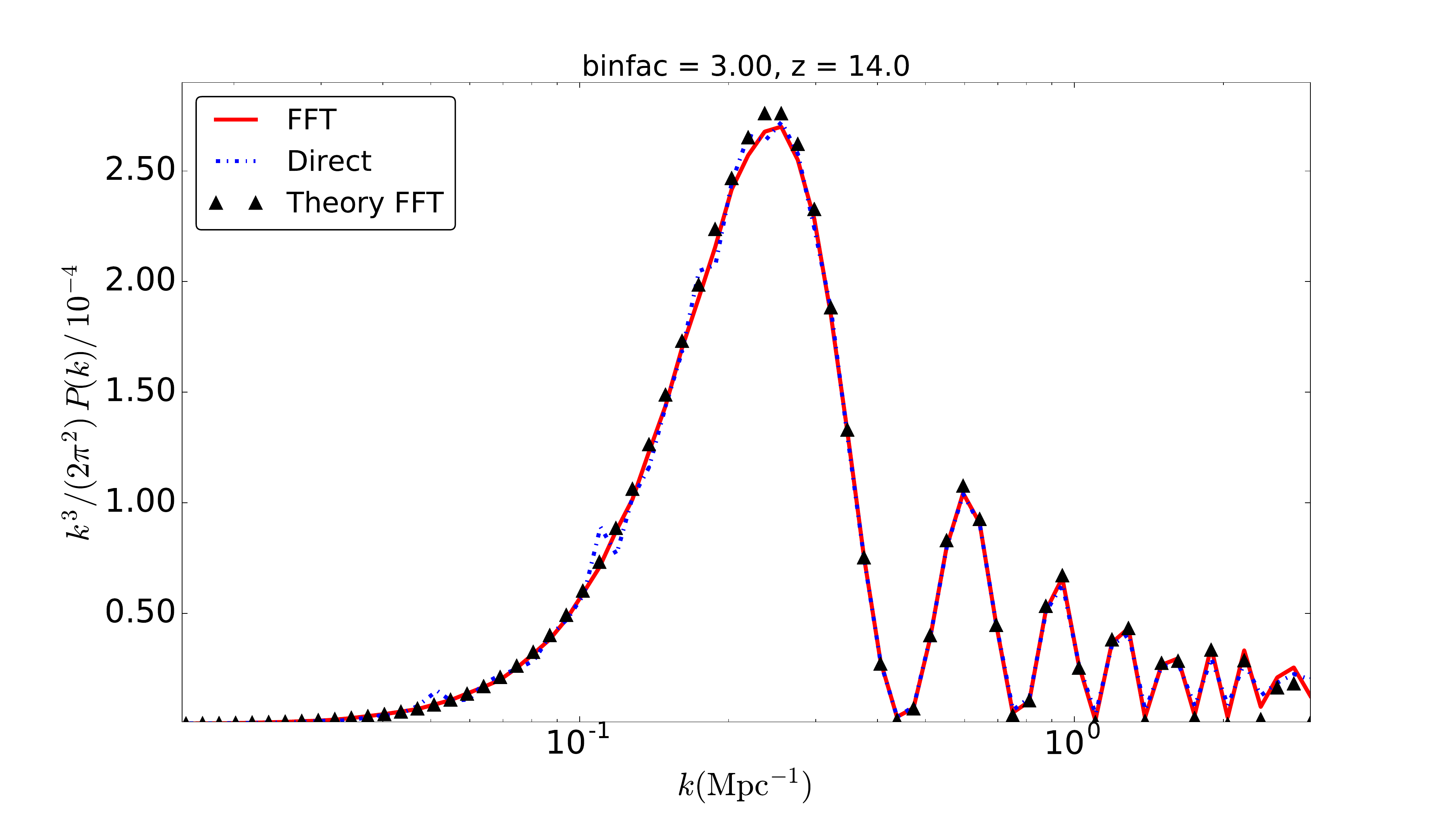}\\
  \end{array}$
  \caption{Spherically-averaged power spectrum from a simulation of reionization
  that assumes the ionization field consisting of randomly-distributed uniform-sized
  spherical bubbles.
  The FFT estimator, direct method and theory are perfectly in agreement for the
  spherically-averaged power spectrum with only slight divergence at very small
  scales, caused by overlap of ionized spheres being allowed in the simulations,
  but not in theory.
  The box analysed here have $z=14$ and $\xh = 0.99$ (chosen to minimize
  differences between simulation and theory due to overlap).
  }
  \label{fig:powerSpec}
\end{figure}

We analyse ionization boxes with 600 pixels and 600 Mpc on a side,
because this is the resolution of the simulations of \citet{Watkinson2015a}
from which we ultimately wish to study the bispectrum during the cosmic dawn
and the EoR in future work.
We arbitrarily set the radius of the bubbles to be 10 Mpc,
choosing smaller bubbles to minimize the effect of overlap.
Unlike the density simulations, the power spectrum from the model of \citet{Bharadwaj2005}
is not monotonic in $k$.
We therefore use this to test the FFT estimator for the power spectrum,
i.e. Equation \ref{eq:Jeong_PS}.\footnote{We again emphasize that anyone
just interested in the spherically-averaged power spectrum should stick with the
standard direct-measurement method, as in this case it is faster than the FFT estimator.}
For direct estimation of the power spectrum we loop through the FT box and calculate
$\langle \delta(k)^2\rangle$ for all $k$ that fall in a given bin.
In Figure \ref{fig:powerSpec}, we plot the spherically-averaged power spectrum
normalised by $k^3/2\pi^2$, i.e. the dimensionless power spectrum.
We find that there is good agreement between the FFT and direct methods,
as well as with the theoretical prediction of Equation \ref{eqn:BharPk}.

\begin{figure*}
  \begin{minipage}{176mm}
  \begin{tabular}{c}
    \includegraphics[trim = 0.9cm 2.65cm 3.0cm 2.1cm, clip=true, scale=0.25]{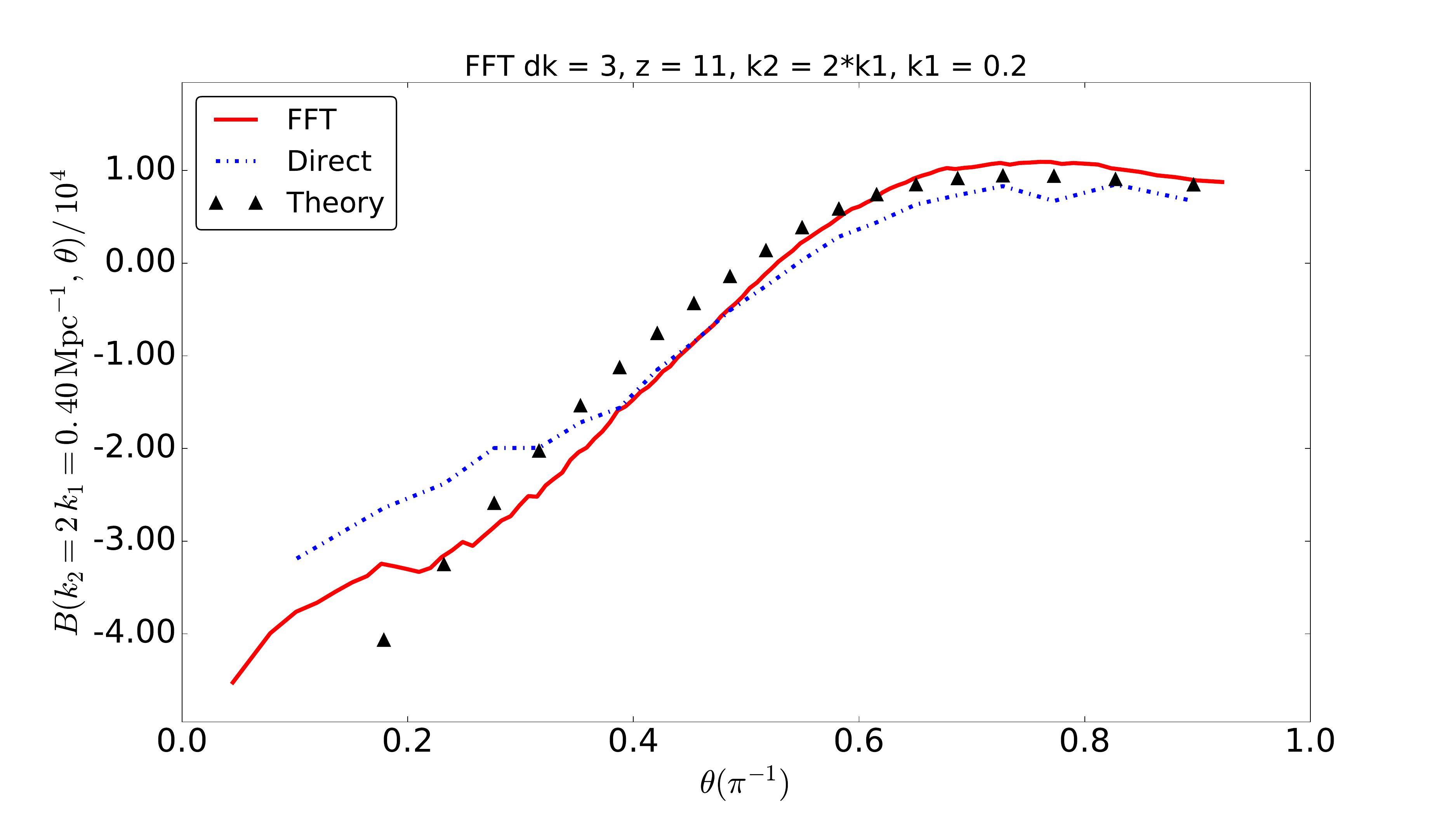}\includegraphics[trim = 1.3cm 2.65cm 0.5cm 2.1cm, clip=true, scale=0.25]{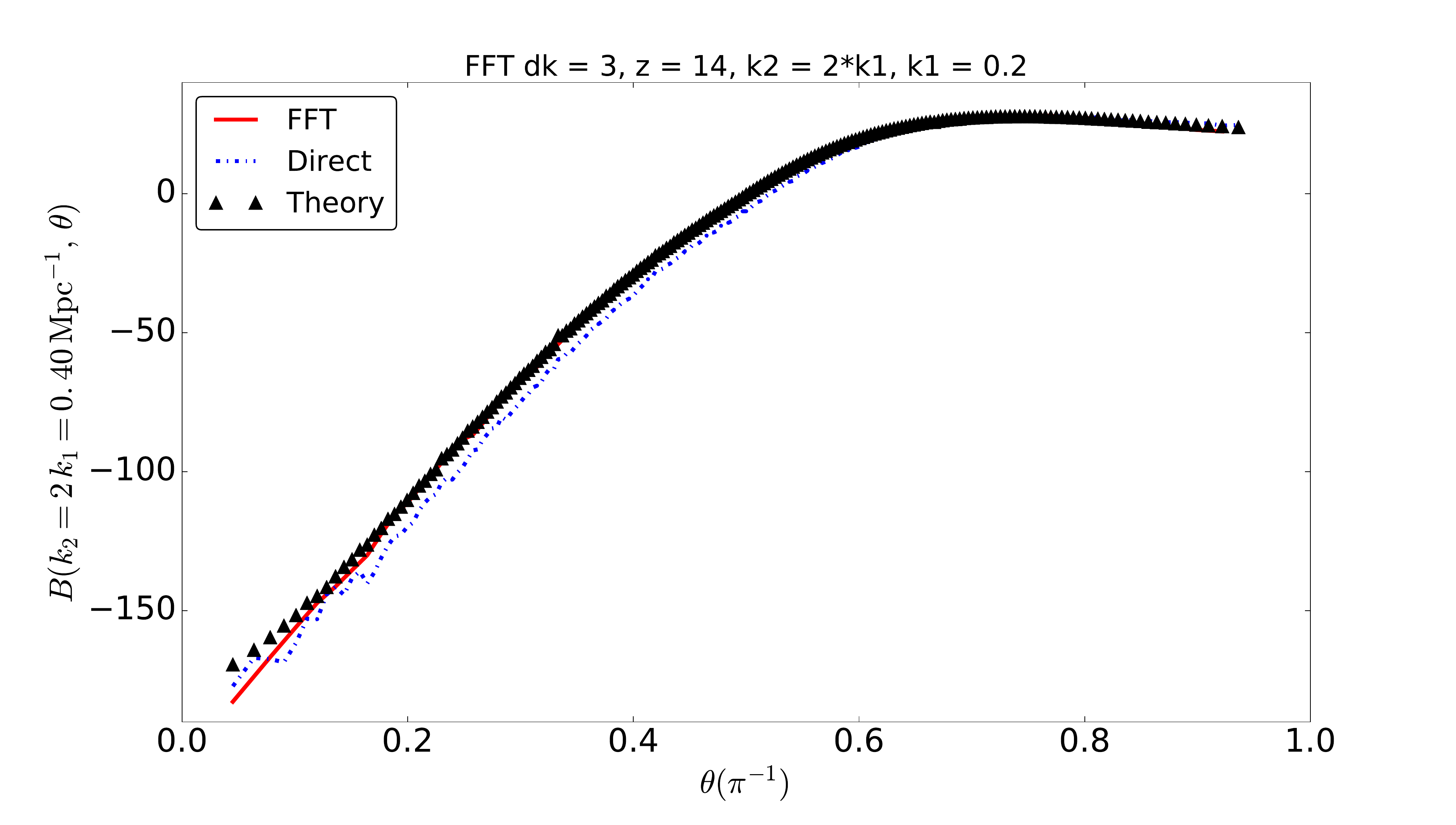}\\
    \includegraphics[trim = 0.9cm 2.65cm 3.0cm 2.1cm, clip=true, scale=0.25]{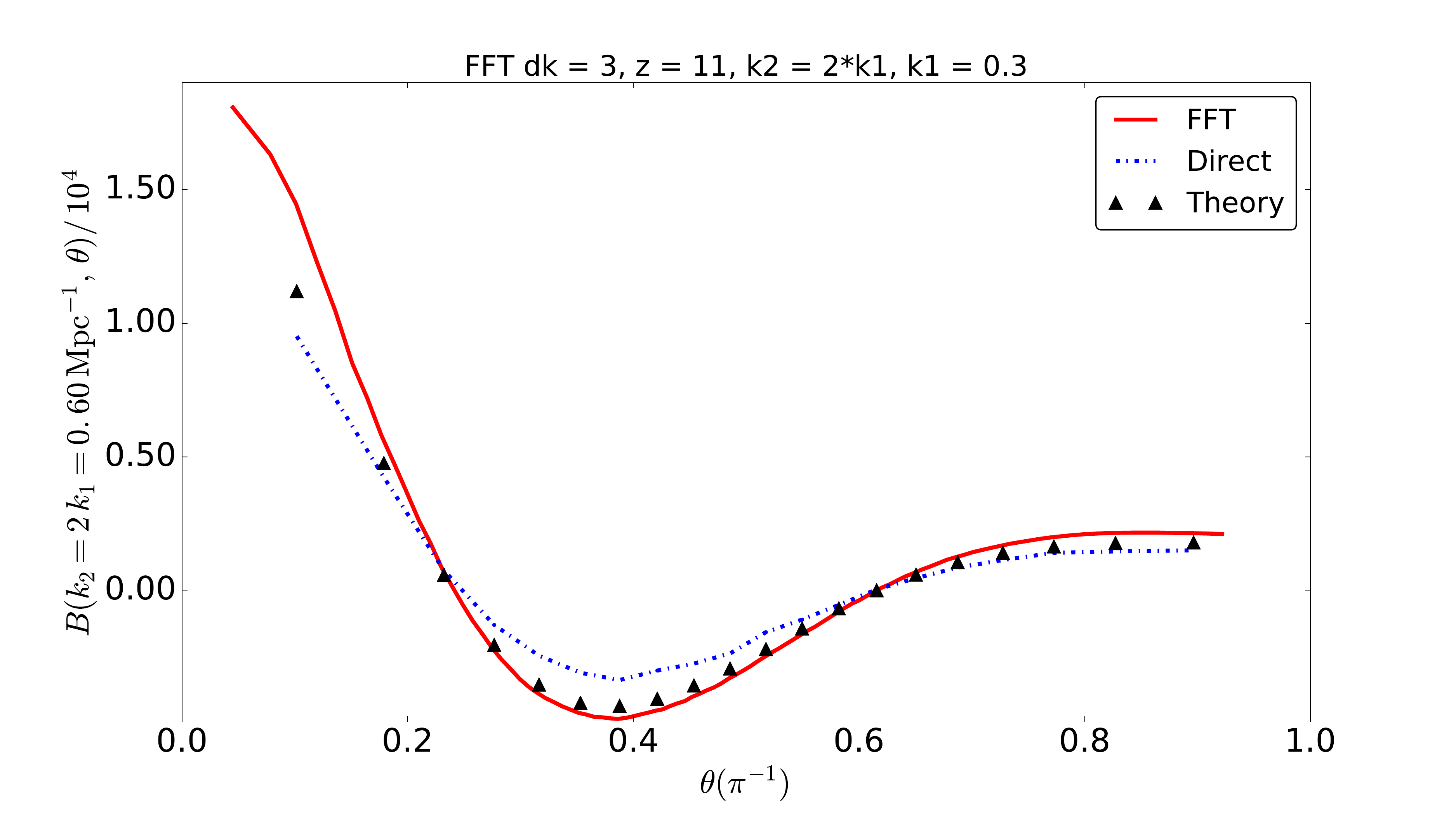}\includegraphics[trim = 1.3cm 2.65cm 0.5cm 2.1cm, clip=true, scale=0.25]{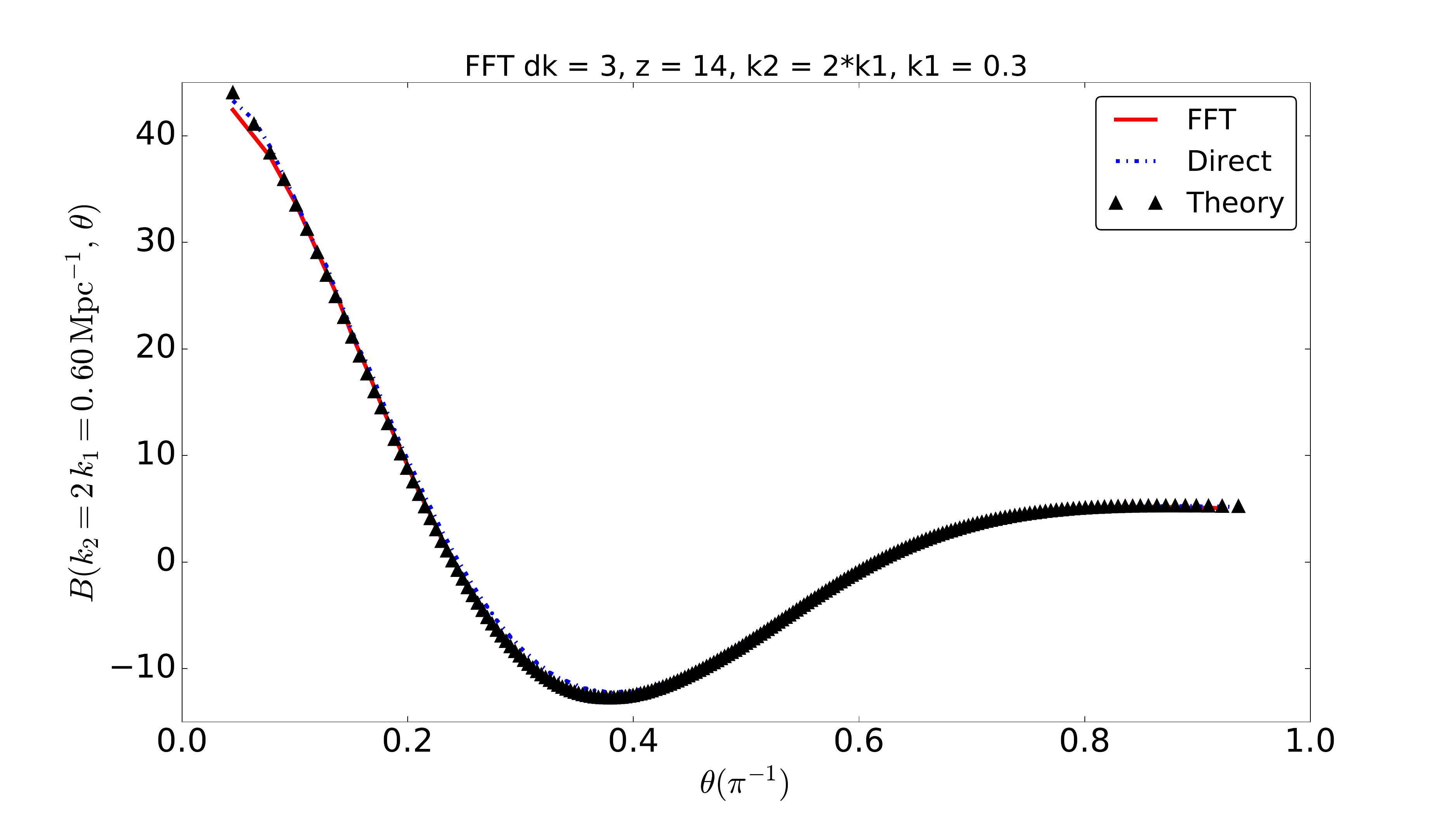}\\
    \includegraphics[trim = 0.9cm 0.5cm 3.0cm 2.1cm, clip=true, scale=0.25]{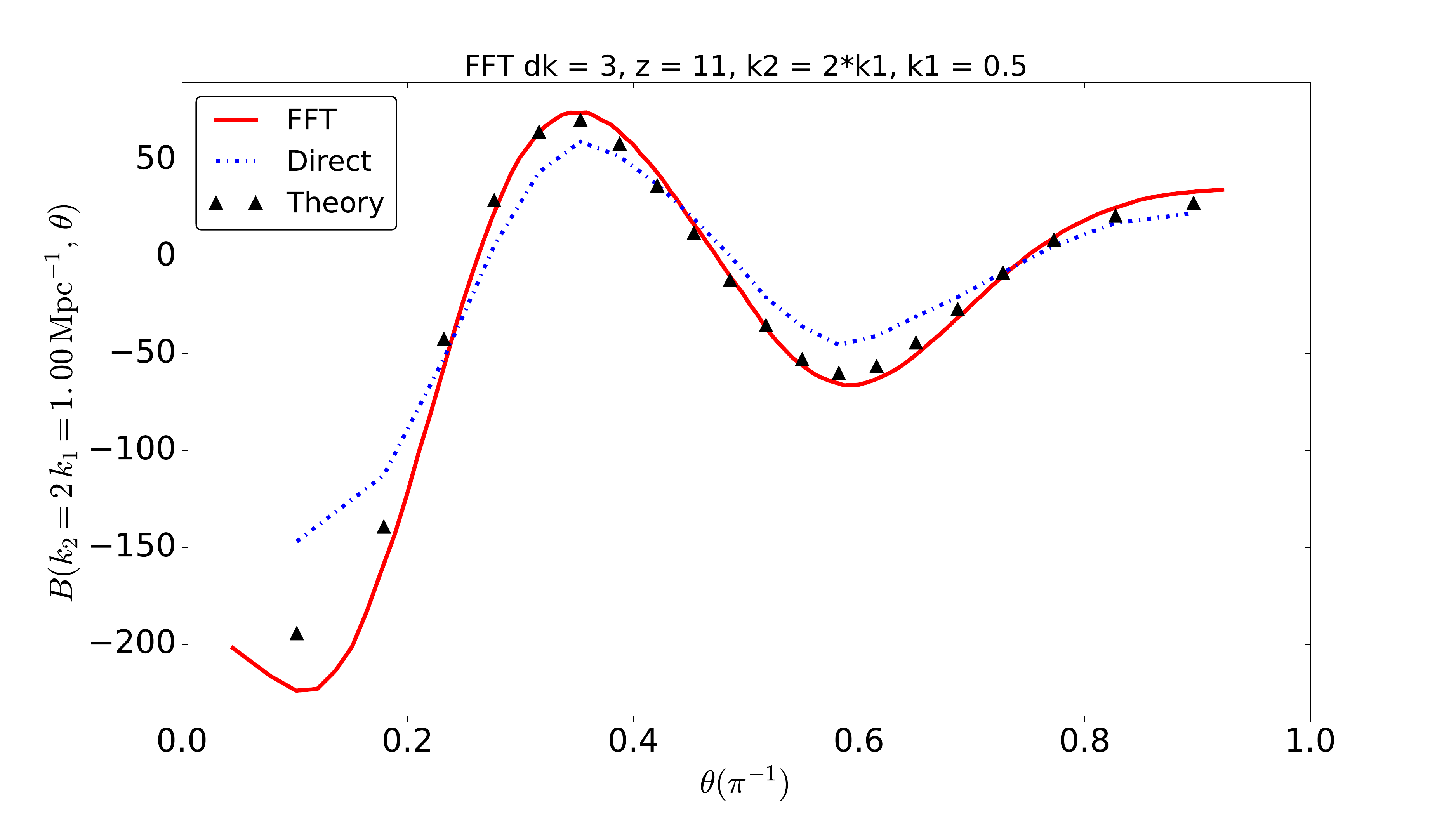}\includegraphics[trim = 1.3cm 0.5cm 0.5cm 2.1cm, clip=true, scale=0.25]{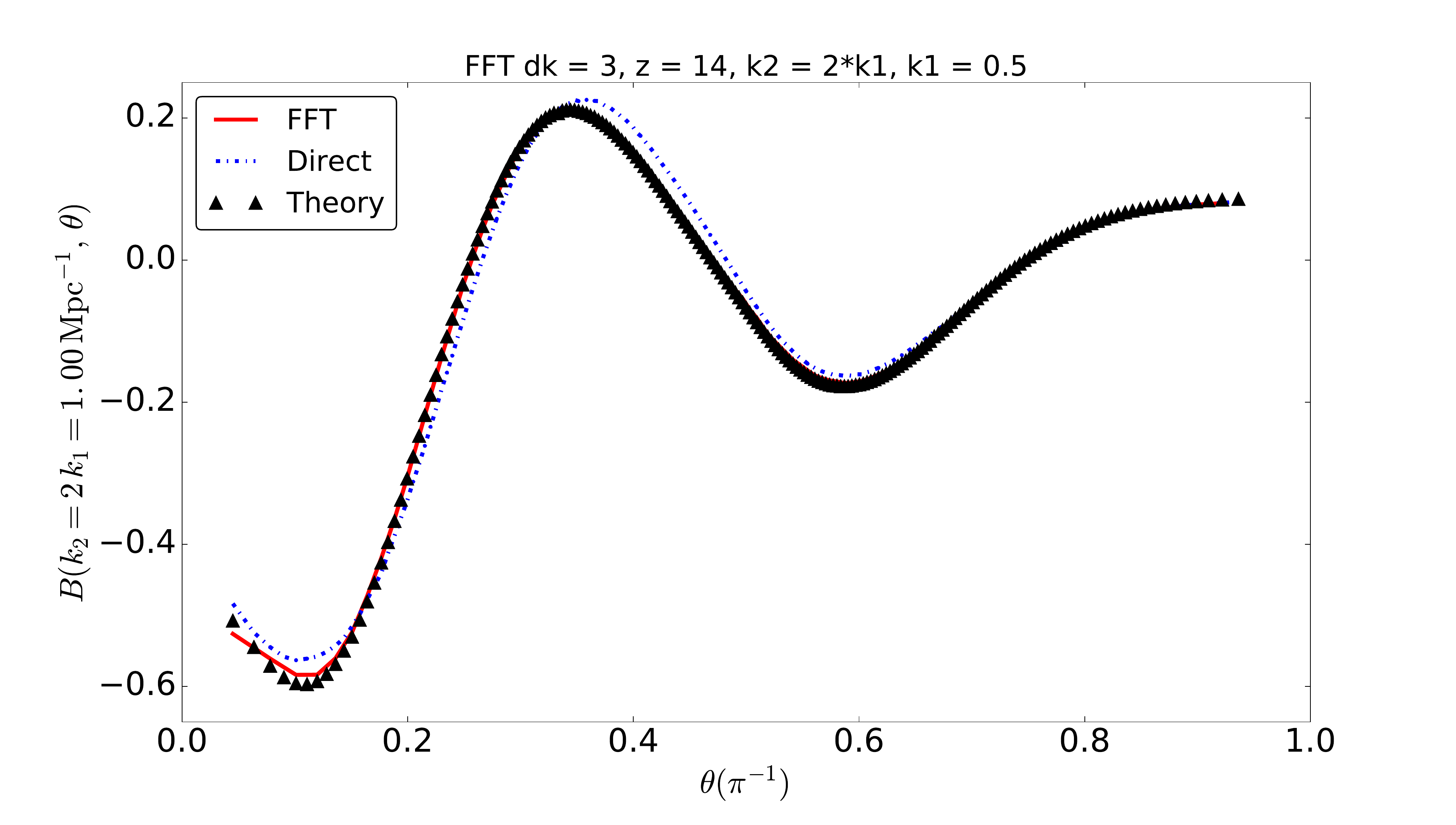}\\
    \includegraphics[trim = 0.9cm 0.3cm 3.0cm 2.1cm, clip=true, scale=0.25]{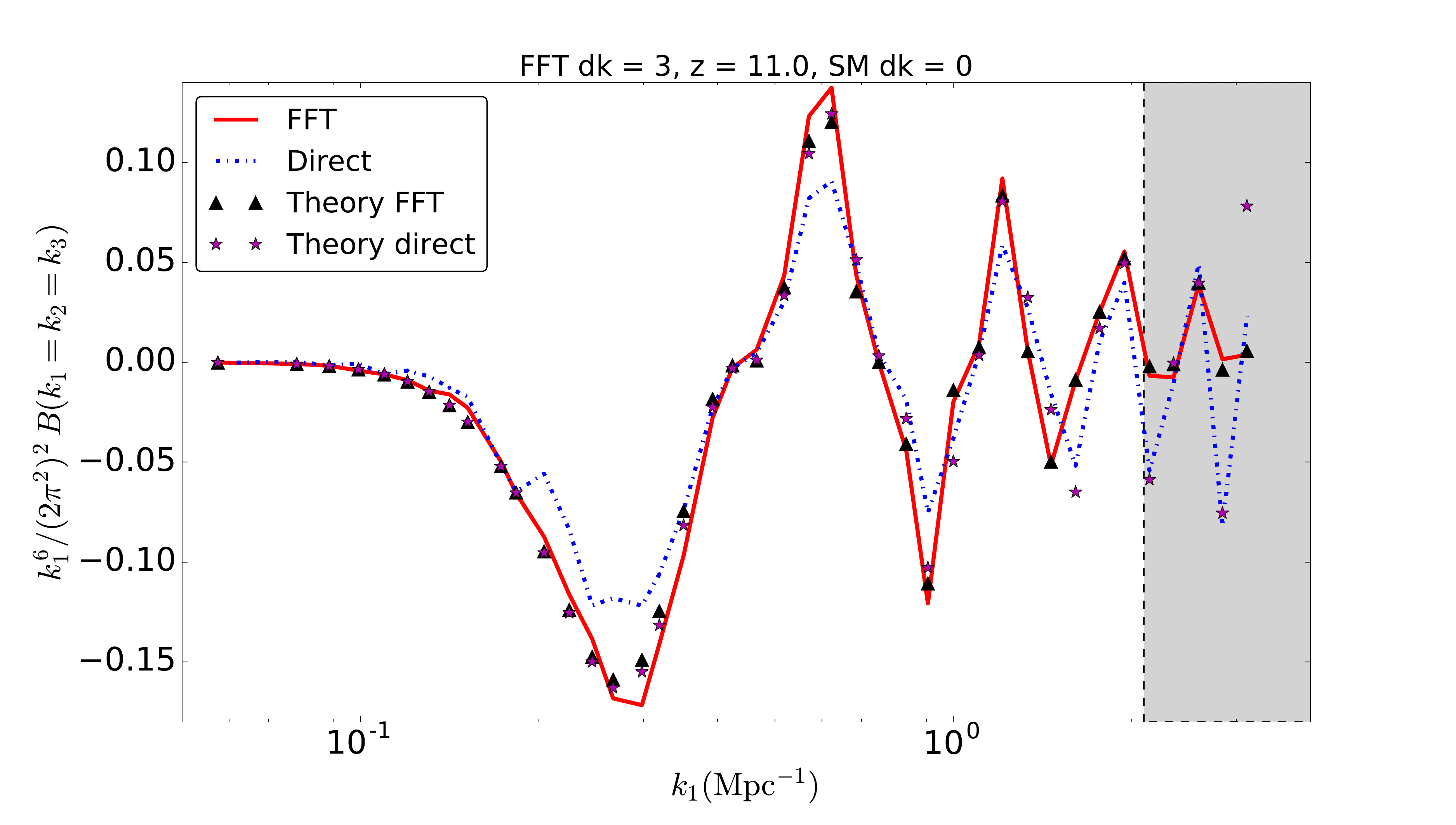} \includegraphics[trim = 1.3cm 0.3cm 0.5cm 2.1cm, clip=true, scale=0.25]{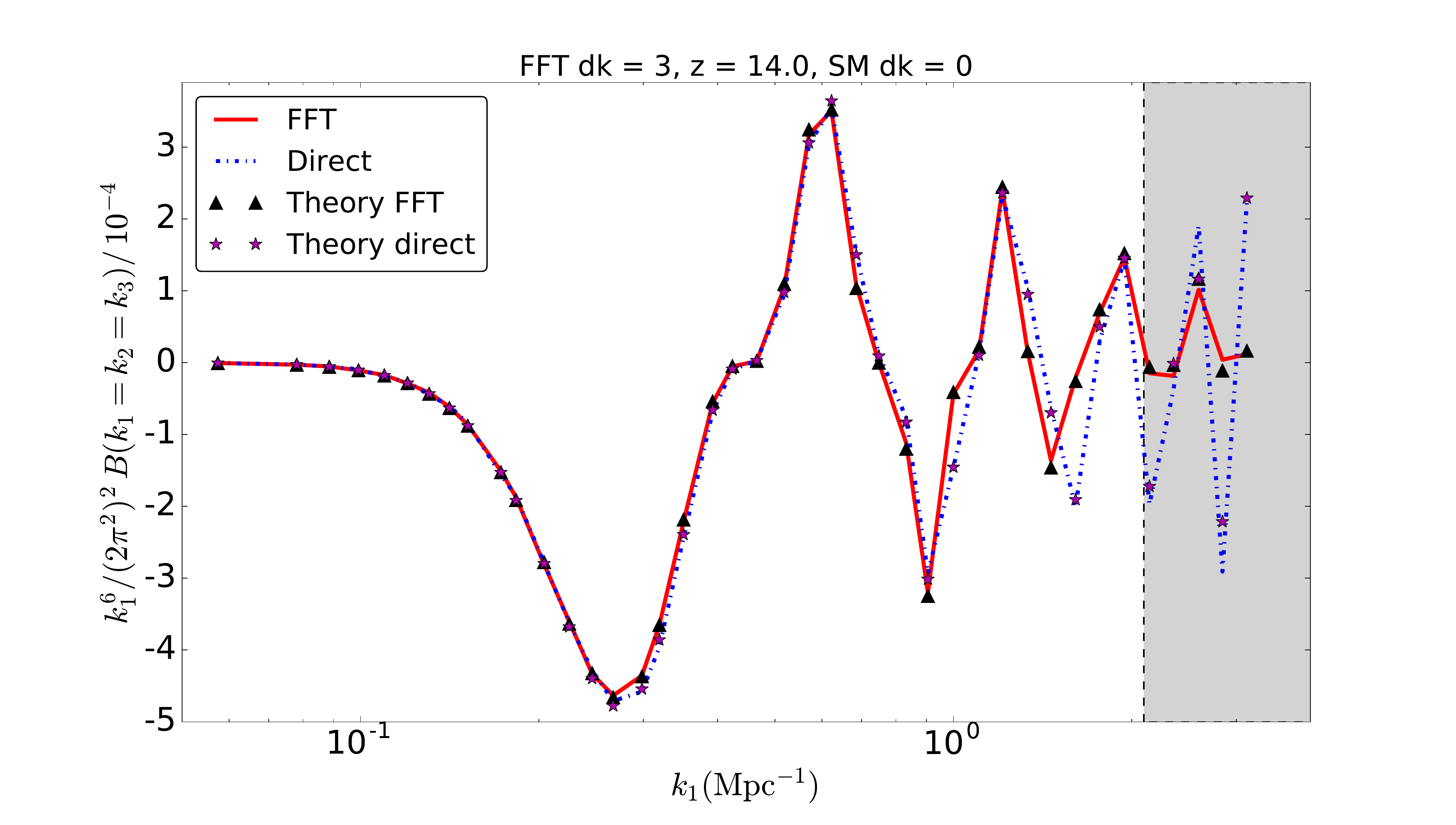}\\
  \end{tabular}
  \caption{Bispectrum from a simulation of reionization that assumes
  the ionization field is made up of randomly-distributed uniform-sized spherical
  bubbles.
  Pink stars mark the theoretical prediction as calculated using the same
  binning as the direct method.
  We plot configurations where $k_2=2\,k_1$ and
  $k_1 = 0.2\,\mathrm{Mpc}^{-1}$, $k_2 = 0.4\,\mathrm{Mpc}^{-1}$ (top),
  $k_1 = 0.3\,\mathrm{Mpc}^{-1}$, $k_2 = 0.6\,\mathrm{Mpc}^{-1}$ (middle-top),
  $k_1 = 0.5\,\mathrm{Mpc}^{-1}$, $k_2 = 1.0\,\mathrm{Mpc}^{-1}$ (middle-bottom),
  and the equilateral configuration (bottom).
  The evolution of the ionization field corresponds to $z=11$
  and $\xh = 0.88$ (left column) ,and $z=14$ and $\xh = 0.99$ (right column).
  The grey shaded area corresponds to $k$ values beyond which
  \citet{Sefusatti2015}
  predict that the FFT bispectrum estimator will become inaccurate.
  As with the density field, the FFT estimator performs well,
  as differences between the two methods can be attributed to binning.
  This is underlined by how well each method follows the theory when calculated
  with the same binning scheme in the right column (at this stage overlap
  will be minimal and so differences between measurements from the simulation
  and theory due to overlap will be minimized).
  }
  \label{fig:xH_bi}
  \end{minipage}
\end{figure*}

In Figure \ref{fig:xH_bi}, we plot the bispectrum from the reionization simulation
as a function of $\theta$ with $k_2 = 2\,k_1$ for $k_1 = (0.2, 0.3, 0.5)\,\mathrm{Mpc}^{-1}$
and $k_2 = (0.4, 0.6, 1.0)\,\mathrm{Mpc}^{-1}$
from top to middle-bottom respectively.
The equilateral configuration is shown in the bottom row of Figure \ref{fig:xH_bi},
here we normalise the bispectrum by $k^6/(2\pi^2)^2$ to highlight the
oscillatory nature of the signal.
We bin the direct estimates of the bispectrum with $\cos(\theta)\pm0.02$.
The left column corresponds to $z=11$ when the neutral fraction is 0.88,
and the right column to $z=14$ when the neutral fraction is 0.99.

There are a few interesting features of the bispectrum for this model,
which are most clear in plots of the bispectrum for the equilateral configuration
(see bottom row of Figure \ref{fig:xH_bi}).
As is to be expected there is a main peak around the $k$ associated with the bubble size,
i.e. $k=2\pi/R$.
Following this peak is a ringing due to the spheres having hard edges.
There is also a negative minimum, in the normalised bispectrum, around the
scale associated with twice the bubble size, this occurs because the unnormalised
bispectrum plateaus towards a constant negative value with decreasing k,
and then the signal is suppressed towards zero by the normalisation.
Such features are defined by the window function and
vary only in amplitude as the ionised fraction increases do to the presence of
more spherical ionised bubbles.

We see that the bispectrum as measured by the FFT-bispectrum estimator follows
the theoretical predictions very closely. 
Again the bispectrum becomes noisy due to sample variance for $k\lesssim (100/6)\,k_{\mathrm{f}}$,
which for this dataset corresponds to $k\le 0.17$ Mpc$^{-1}$.
This is only evident in the unnormalised bispectrum, which we do not show here,
and is far less pronounced for our ionization field than it is with the density field.
For example, we do not see the sign of the bispectrum switching from negative to positive
in this regime, as we do for the density field.
It is likely that this is because our reionisation simulations are very simple;
the ionization field is binary and so they will contain very little numerical noise
as compared to the density field.

We again find that the estimator follows the theoretical predictions very closely
where $k\ge k_{\mathrm{max}}$, this is most plain to see from the plots of
equilateral configurations in the bottom row of Figure \ref{fig:xH_bi}.
For the $z=11$ model, we do see a slight divergence from theory at certain
values of $k$.
This is clearly due to the allowance of bubble overlap in the simulation,
as can be seen by comparing the left and right columns of the bottom row of
Figure \ref{fig:xH_bi}.
For example, we see that the FFT bispectrum estimator starts to diverge
slightly from theory at $k<0.2$ at $z=11$, where $\xh\sim 0.9$,
whereas it follows the theoretical predictions very closely when $z=14$,
where $\xh\sim 0.99$.

\subsection{Effect of binning in the direct estimator of bispectrum}
As discussed in Section \ref{sec:direct_est},
there is a clear difference between the binning
approaches of the two bispectrum algorithms discussed in this paper.
When the bispectrum is an oscillatory function of $k$, increases to the
$k$ bin width will cause the measured bispectrum to diverge
dramatically from its true value.
The bispectrum of the toy model, introduced in Section \ref{sec:toy_model},
is a perfect example of such a scenario.
The behaviour of the bispectrum in
this model is determined by the window function $W(kR)$, as shown in
Equation \ref{eqn:BharBk}, which is very oscillatory in
nature.
To reduce sample variance at small $k$ values
for equilateral triangle configuration (i.e. $k_1 = k_2 =k_3$),
if one increases the $k_1$ bin width significantly, one would
essentially vary the $W(k_1R)$ function in the bin. If some of the
$k_1$ values within the bin lie somewhere close to the dips of
oscillations in $W(k_1R)$, the bispectrum estimation by different
triangles contributing within bin will vary severely, as the change in
amplitude near the dips of oscillation is large.
Thus, the bin-averaged bispectrum in such a scenario will differ
significantly from the theoretical expectation value of the bispectrum,
as predicted by Equation \ref{eqn:BharBk}, and as estimated using the
mid-point or the average value of the $k$ mode in the respective bin. To
avoid this we keep the $k$ bin width at its bare minimum.
and thus the direct method's bispectrum estimation for small $k$ values
is more affected by sample variance for this toy model
(see Figure \ref{fig:xH_bi}).

For other triangle configurations, where $k_1 \neq k_2 \neq k_3$, the
situation would be a bit more complicated, as each of the window
function contributing to the bispectrum for that triangle will probe
different parts of this oscillatory window function and their product
will give rise to ``beats''. Different triangles within the same
bispectrum estimation bin will thus produce different beats for the
oscillatory window function, and their average value across the bin
will be very different than the theoretical prediction for the mid
point of the bin.
To demonstrate this point more clearly, we estimate
the theoretical bispectrum, following Equation \ref{eqn:BharBk},
for each of the triangles contributing within a bin, and plot the
bin-averaged theoretical value,
this is shown by the pink stars in Figure \ref{fig:xH_bi}.
We observe that the bin averaged theoretical prediction follows
the numerical estimation very closely.

This discussion makes it clear that great care must be taken when using our direct
method to measure highly oscillatory bispectrum signals, such as that of the toy model for
the ionization field explored here.
However, this toy
model is very limited in nature as it assumes all ionized regions in
the IGM to be spheres of equal radius $R$, at every stage of the
EoR. In reality, the ionized regions, at any stage of
reionization, will be of different shapes and volumes.
This has been observed by various reionization simulations to date
(e.g. \citealt{Majumdar2014,Iliev2015}).
If we consider that at any redshift during reionzation the size of the
ionized spheres are uniformly distributed between
$R_{\mathrm{min}} \leq R \leq R_{\mathrm{max}}$,
and so the resulting 21-cm signal will be proportional to $\sum_i W(kR_i)$.
It can be shown that even for a moderate range of values of $R_i$,
unlike $W(kR)$, $\sum_i W(kR_i)$ is a smooth function of $k$.
Thus, it will be safe to use the direct
estimator of bispectrum in such a scenario. We discuss this in more
details in our follow up work Majumdar et al. (in prep).

\section{Conclusion} \label{sec:Conc}
In this paper we have presented the derivation of a fast estimator for the polyspectra.
We outline an algorithm that provides a further speed up by initialising an
indexing array in which each $j$ index contains an array of all FFT-box co-ordinates
that correspond to $k$-vectors of a particular length
(connected to the array element index $j$ by a scaling factor).
This removes the need to fully loop through the FFT-box for every bispectrum call.

As we intend to apply this approach to study the bispectrum of the 21-cm signal,
we focus our tests of this algorithm on the bispectrum.
We test this FFT-bispectrum algorithm for the bispectrum using a non-linear N-body density field
(a mildly non-Gaussian dataset), and a toy model for reionization consisting of mono-sized ionized spheres.
For both cases our FFT-bispectrum algorithm reproduces the bispectrum predicted by theory
and measured using a direct-measurement algorithm.
We find that the algorithm behaves reasonably well in both test cases at $k>k_{\mathrm{max}}$,
where it has previously been argued that the estimator should break down.
The argument is that a periodicity in the phase term
of the Kronecker-delta function (when enforced using FFTs)
will cause the estimator to become inaccurate beyond $k_{\mathrm{max}}$.
We argue that the reason we do not see the estimator break down, is because
the impact of this periodicity will be within the magnitude of inaccuracy introduced
by using FFTs to enforce the Kronecker delta, which is inherent to the estimator at
all $k$.
This inaccuracy occurs as the contribution from the Kronecker-delta term is not
exactly unity,
this is because discretised $k$ vectors often do not form perfectly closed triangles.
We therefore suggest that the FFT-bispectrum estimator may still be applied
in this regime.
We also advocate using a bin width of just one pixel when measuring the bispectrum with
the FFT estimator, and applying any desired binning subsequently.

At low $k$, both estimators become noisy due to sample variance, and this can
cause erratic behaviour, including the sign of the bispectrum randomly flipping
from negative to positive, and vice versa.
This erratic behaviour can be suppressed by plotting the amplitude of the bispectrum,
with the drawback of suppressing genuine sign changes in the signal,
which may contain important information.

The FFT-polyspectra algorithm presented in this paper is faster than
direct-measurement methods and is fast enough to be used in sampling problems.
Given the non-Gaussianity of the 21-cm signal during the cosmic dawn and reionization,
this estimator will be invaluable for performing parameter estimation.
Furthermore, whilst we focus on cosmological datasets, this algorithm
will be very valuable for any non-Gaussian dataset.

\section*{Acknowledgements}
The authors would like to thank Emiliano Sefusatti, Donghui Jeong, Keri Dixon,
Ilian Iliev and
Adam Lidz for their insightful comments.
CW thanks the Science and Technology Facilities
Council via the SKA-preconstruction-phase-continuation grant. SM
and JRP acknowledge support under FP7-PEOPLE-2012-CIG grant
\#321933-21ALPHA, and the European Research Council under ERC grant
number 638743-FIRSTDAWN.

\bibliographystyle{mn2e}


\bsp
\end{document}